\documentclass[aps,pre, onecolumn,12pt, notitlepage, showpacs,floatfix,nofootinbib, double-space,superscriptaddress]{revtex4-2}

\usepackage{subcaption}
\usepackage{bbold}
\usepackage{amssymb}
\usepackage{amsfonts}
\usepackage{amsmath}
\usepackage{amsthm}
\usepackage{epsfig}
\usepackage{float}
\usepackage{multirow}
\usepackage[usenames,dvipsnames]{color}
\usepackage[latin1]{inputenc}
\usepackage{xr}
\usepackage{enumitem}
\usepackage{microtype}
\usepackage{graphics,graphicx,xcolor}
\usepackage{comment}
\usepackage{cancel}
\usepackage{lipsum}

\newcommand{\av}[1]{\left \langle #1 \right \rangle}
\newcommand{\sig}[2]{\vec{\sigma}_{#1}^{#2}}
\newcommand{\kk}[1]{\mathbf{K}_{#1}}

\graphicspath{{newfigs/}}

\begin{document}

\title{Arnold tongues in the forced Kuramoto model with matrix coupling}

\author{Guilherme S. Costa}
\affiliation{Instituto de F\'isica "Gleb Wataghin", Universidade Estadual de Campinas, Unicamp 13083-970, Campinas, SP, Brazil}

\author{Marcus A. M. de Aguiar}

\affiliation{ICTP South American Institute for Fundamental Research \& Instituto de F\'isica Te\'orica - UNESP, 01140-070, S\~ao Paulo, Brazil }
\affiliation{Instituto de F\'isica "Gleb Wataghin", Universidade Estadual de Campinas, Unicamp 13083-970, Campinas, SP, Brazil}
	
\begin{abstract}

We consider a generalization of the Kuramoto model in which phase oscillators are represented by unit vectors coupled by a matrix of constant coefficients. We show that, when the oscillators are driven by an external periodic force, several resonances appear, giving rise to Arnold tongues that can be observed as the intensity and frequency of the external force are varied. Applying the Ott-Antonsen ansatz we obtain equations for the module and phase of order parameter. As these equations are explicitly time-dependent, we resort to extensive numerical simulations to uncover the resonant modes and their associated Arnold tongues and devil's staircases. These results contrast with the original forced Kuramoto model, where only $1:1$ resonance is possible.

\end{abstract}

\maketitle

\section{INTRODUCTION}

Synchronization and collective behavior are present in a myriad of dynamic phenomena \cite{pikovsky2001synchronization,boccaletti2018synchronization}. From pacemaker heart cells \cite{Michaels704} to lasers \cite{Xiuqi2023}, these systems are composed of several elements that operate collectively to perform specific functions. The synchronization problem was placed on a solid mathematical footing by Winfree \cite{Winfree1967} and Kuramoto \cite{kuramoto1975}, and since then, several advances have been made in the field, considering more complex models and more general types of interactions \cite{Skardal2020,Chandrasekar2020,Lizarraga2023,Costa2025}. A particularly important form of synchronization is {\it entrainment}, which occurs when the system adjusts its behavior according to external signals, such as circadian rhythms, where changes in temperature and light exposition control the dynamics of biological oscillators \cite{Murayama2017,Prokkola2018,Jimenez2022,Sanchez2022}. Other examples of entrainment can be found is nanomechanical oscillators \cite{Shim2007}, detonation in gaseous mixtures \cite{Kasimov2022} and quantum phase localization \cite{Ali2024}.

The quintessential model for investigating synchronization problems is the Kuramoto model \cite{kuramoto1975}, in which $N$ oscillators, each described by a single phase $\theta_i$ and natural frequency $\omega_i$, interact with all other oscillators following the dynamical equation
\begin{equation}
    \dot{\theta_i} = \omega_i + \dfrac{K}{N} \sum \sin (\theta_j - \theta_i),
    \label{eq::originalKura}
\end{equation}
where $K$ is a coupling constant. In the limit $N \to \infty$ the system exhibits a continuous phase transition from disorder to collective motion (synchronization) as the coupling increases. For $K$ smaller than a threshold value $K_c$, that depends on the distribution of natural frequencies, the oscillators move as if they were independent  while for $K > K_c$ they start to cluster and eventually operate as a single rotating element as $K \to \infty$. 

Several extensions and generalizations of the Kuramoto model were proposed since its inception, such as the introduction of frustration \cite{Sakaguchi1986}, different types of coupling functions \cite{hong2011kuramoto,yeung1999time,breakspear2010generative} and networks of connections \cite{strogatz2001exploring,moreno2004synchronization,Rodrigues2016},
different distributions of natural frequencies \cite{martens2009exact,Gomez-Gardenes2011,Ji2013}, inertial terms
\cite{Acebron2005,dorfler2011critical,olmi2014hysteretic}, external periodic driving forces 
\cite{Childs2008,moreira2019}, higher dimensions \cite{chandra2019continuous,barioni2021,de2023generalized}, higher order interactions \cite{battiston2020networks,Skardal2020,muolo2024phase,Costa2025} and mobile oscillators \cite{o2017oscillators,sar2026interplay}. 

Of particular importance to this work is a recent generalization of the Kuramoto model that promotes the scalar coupling constant $K$ to a coupling matrix \textbf{K} \cite{Buzanello2022}. This is implemented by using the vector representation introduced in \cite{chandra2019continuous}, where the oscillator's phase $\theta_i$ is replaced by the unit vector $\vec\sigma_i = (\cos\theta_i,\sin\theta_i)$. Coupling these unit vectors by a matrix breaks the rotational symmetry and introduces new synchronized phases, such as active and phase tuned states. 

In this work, we investigate the effects of external periodic forces on the matrix coupled Kuramoto model. We show that the combination of symmetry breaking and periodic forcing leads to the appearance of resonances and Arnold tongues. We show that the Ott-Antonsen ansatz \cite{Ott2008} can be used to find dynamical equations for the order parameter but, in contrast with the original forced Kuramoto model \cite{Childs2008}, the symmetry breaking introduced by the matrix coupling prevents the elimination of the time-dependence by a change of reference frame. As a consequence the dynamical equations are explicitly time-dependent and analytical solutions become impracticable. Using numerical integration we uncover several entrainment modes in the form of Arnold tongues \cite{Arnold2009}, not present in the original model.

We show that when the unforced system is in an oscillatory state, multiple higher order resonances and Arnold tongues appear in the phase space. In phase tuned states, on the other hand, the external forcing induces only relatively simple resonances, and the phase space is still dominated by phase tuned modes and $1:1$ locked states. 

This paper is organized as follows: in Section II, we revisit the matrix coupled Kuramoto model, discuss connections with other synchronization models and add an external periodic force acting on the oscillators. We also perform a dimensional reduction using the Ott-Antonsen ansatz and derive the dynamical equations for the order parameter. In Section III, we review the different ways to characterize the entrainment between external forces and the system.  We discuss and characterize the dynamical behavior of the model in Section IV, focusing on the two distinct behaviors induced by the frustration: oscillatory and phase tuned states. Finally, we discuss our results in Section V, where we argue that the symmetric part of the coupling matrix can be interpreted as an internal force that resonates with the external one to give rise to the observed tongues.

\section{MATRIX COUPLING AND EXTERNAL FORCES}
\label{sec::modMatrixCoup}

In this section we introduce the matrix coupled Kuramoto model, initially proposed in \cite{Buzanello2022}. Following the notation introduced in \cite{chandra2019}, instead of describing the oscillators by their phases $\theta_i$, we use unit vectors $\vec{\sigma_i} = (\cos \theta_i,\sin \theta_i)$. Using Eq.(\ref{eq::originalKura}) it is not hard to show that the vectors satisfy the equation 
\begin{equation}
	\label{eq::vecform}
	\dfrac{d \sig{i}{}}{dt} = \textbf{W}_i \sig{i}{} + \dfrac{K}{N} \sum_{j=1}^{N} [ \sig{j}{} - (\sig{i}{}\cdot \sig{j}{}) \sig{i}{}],
\end{equation}
in which 
\begin{equation}
	\textbf{W}_i = 
	\begin{pmatrix} 
		0 & \omega_i \\
		-\omega_i &0 
	\end{pmatrix}
\end{equation}
is the frequency matrix. Although Eq. (\ref{eq::vecform}) is completely equivalent to Eq.(\ref{eq::originalKura}), the vector formulation has the advantage of being easily extended to higher dimensions by simply interpreting $\vec\sigma_i$ as $D-$dimensional vectors, described by $D-1$ spherical angles, that rotate on the surface of unit $D-$spheres. The only requirement is that frequency matrix $\textbf{W}_i$ must be anti-symmetric \cite{chandra2019}.

In this formulation, the order parameter can be calculated as the average of the vectors $\sig{i}{}$
\begin{equation}
	\vec{r} = \dfrac{1}{N}\sum_{i=1}^{N} \vec{\sigma}_i  = (r \sin{\psi}, r\cos\psi).
	\label{eq::orderP}
\end{equation}
Another important advantage of the vector formulation is that it leads to the natural extension of the coupling constant $K$ to a $D \times D$ real matrix $\kk{}$ \cite{Buzanello2022}, that acts on the vectors and modifies the dynamics according to
\begin{equation}
	\dfrac{d \sig{i}{}}{dt} = \textbf{W}_i \sig{i}{} + \dfrac{1}{N} \sum_{j=1}^{N} [ \mathbf{K}\sig{j}{} - (\sig{i}{}\cdot \mathbf{K} \sig{j}{}) \sig{i}{}].
	\label{eq::kuraFrust}
\end{equation}

The general interpretation of the coupling matrix is that it plays the role of generalized frustration, in which the vector $\sig{j}{}$ is modified by $\kk{}$ before interacting with $\sig{i}{}$. Using the definition in Eq.(\ref{eq::orderP}), the dynamical equation can be simplified as
\begin{equation}
	\dfrac{d \sig{i}{}}{dt} = \textbf{W}_i \sig{i}{} + [ \mathbf{K}\vec{r} - (\sig{i}{}\cdot \mathbf{K} \vec{r}) \sig{i}{}].
	\label{eq::kuraFrustMF}
\end{equation}

Several versions of the Kuramoto model can be obtained as particular cases of Eq.\eqref{eq::kuraFrustMF} by fixing $D=2$ and choosing the matrix $\kk{}$ accordingly \cite{costa2024dynamics}. For example, if $\kk{}$ is proportional to the identity matrix, $\kk{} = K \mathbb{1}$, we recover the original Kuramoto model, Eq. \eqref{eq::originalKura}. If, instead, $\kk{}$ is proportional to a rotation matrix
\begin{equation}
	\textbf{K} =  K 
	\begin{pmatrix}
		\cos \alpha       & \sin \alpha  \\
		-\sin \alpha & \cos \alpha 
	\end{pmatrix},
\end{equation}
the equation for the phases corresponds to the Kuramoto-Sakaguchi model \cite{Sakaguchi1986}
\begin{equation}
	\dot{\theta}_i = \omega_i + \frac{1}{N} \sum_{j=1}^{N} K \sin (\theta_j - \theta_i - \alpha) .
\end{equation}
Finally, if
\begin{equation}
	\textbf{K}  =  
	\begin{pmatrix}
		0 & 0  \\
		0 & \lambda
	\end{pmatrix}
\end{equation}
the equation for the phases is similar to that proposed by Winfree \cite{Winfree1967}, 
\begin{equation}
	\dot{\theta_i} = \omega_i + \lambda \cos\theta_i \sum_{j=1}^{N} \sin{\theta_j}
\end{equation}  
where $\cos\theta$ is the response function and $\sin\theta$ is the pulse function. Notice that, in this case, rotational symmetry is broken, as $\kk{}$ has real eigenvalues and a single real eigenvector in the y-direction, that provides a special direction in the plane. 

The examples above suggest that, for $D=2$, which is the case of interest in this work, the four independent components of a general coupling matrix $\kk{}$ can be conveniently parameterized by the sum of a rotation (anti-symmetric) matrix $\kk{R}$ and a symmetric matrix $\kk{S}$ as \cite{Buzanello2022}
\begin{equation}
	\textbf{K} \equiv \kk{R} + \kk{S} =  K 
	\begin{pmatrix}
		\cos \alpha       & \sin \alpha  \\
		-\sin \alpha & \cos \alpha 
	\end{pmatrix}
	+ J
	\begin{pmatrix}
		-\cos \beta       & \sin \beta  \\
		\sin \beta & \cos \beta 
	\end{pmatrix},
	\label{matrizK}
\end{equation}
reducing the equations for the phases $\theta_i$ to a familiar form:
\begin{equation}
	\dot{\theta}_i = \omega_i + \sum_{j=1}^{N} \left[ K \sin (\theta_j - \theta_i - \alpha) + J \sin(\theta_j + \theta_i + \beta)\right].
	\label{eq::thetai}
\end{equation}

This equation arises in the context of Stuart-Landau limit-cycle oscillators with conjugate feedback \cite{Chandrasekar2020}. Using Eq.(\ref{eq::orderP}) with $\vec{r}=r(\cos\psi,\sin\psi)$ and defining $\vec{q} \equiv \kk{} \vec{r} \equiv  q (\cos\gamma, \sin \gamma)$, Eq.(\ref{eq::kuraFrustMF}) can also be written as
\begin{equation}
	\dfrac{d \sig{i}{}}{dt} = \textbf{W}_i \sig{i}{} +  [ \vec{q} - (\sig{i}{}\cdot \vec{q}) \sig{i}{}]
	\label{eq::kuraFrust2}
\end{equation}
and Eq.(\ref{eq::thetai}) for the phases becomes
\begin{equation}
	\dot{\theta}_i = \omega_i  +  q \sin(\gamma - \theta_i).
\end{equation}
For $J=\alpha=0$, $\kk{}$ is proportional to the identity, leading to the usual Kuramoto model with $\vec{q} = K \vec{r}$ and $\gamma = \psi$. For $K=0$, on the other hand, identical oscillators with $\omega_i=0$ exhibit full synchronization with zero angular velocity at $\theta_i=-\beta/2$ if $J>0$. These are termed {\it phase tuned} states, as the phase of the cluster can be controlled by $\beta$, which points in the direction of the dominant eigenvector of $\kk{}$. For $K, J \neq 0$, the system also exhibits as active states, where the phase and module of $\vec{r}$ oscillate in time. As rotational symmetry is generally broken, the value of $\omega_0$, the center of distribution of natural frequencies $g(\omega)$, plays a direct role in the dynamics. In \cite{Buzanello2022}, the authors derived the phase diagram of the model, calculating a series of conditions for the system to synchronize in each state. It was shown, in particular, that for every value of $\omega_0$, it is possible to find a range of parameters $K,J,\alpha$ and $\beta$ that reproduces all the possible states of the model.

Finally, following \cite{Childs2008} we introduce an external periodic force acting on the oscillators:
\begin{equation}
		\dot{\theta}_i = \omega_i  +  q \sin(\gamma - \theta_i) + F \sin{(\Omega t - \theta_i)}.
\end{equation}
In the vector formulation this amounts to rewrite Eq.(\ref{eq::thetai}) as
\begin{equation}
	\dfrac{d \sig{i}{}}{dt} = \textbf{W}_i \sig{i}{} +  [ \vec{q} - (\sig{i}{}\cdot \vec{q}) \sig{i}{}] - [ \vec{F} - (\sig{i}{}\cdot \vec{F}) \sig{i}{}]
	\label{eq::kuraFrustForced}
\end{equation}
where $\vec{F} = (F\cos{\Omega t}, F \sin \Omega t)$. In the next subsection we will take the limit of infinitely many oscillators and use the Ott-Antonsen ansatz \cite{Ott2008} to characterize the behavior of the system through its order parameter.

\subsection{Ott-Antonsen ansatz}

In the thermodynamic limit ($N \to \infty$), we define $f(\omega,\theta,t)$ as the density of oscillators at position $\theta$, in time $t$ with natural frequency $\omega$. Since the number of oscillators is conserved, $f$ satisfies the continuity equation
\begin{equation}
	\dfrac{\partial f}{\partial t} + \dfrac{\partial (f v_{\theta})}{\partial \theta} = 0,
    \label{eq::cont}
\end{equation}
with velocity field
\begin{equation}
	\label{eq::velocity}
	v_\theta = \omega + q \sin(\gamma - \theta) + F \sin(\Omega t - \theta) = \omega + \dfrac{1}{2i} \left( He^{-i\theta} - H^*e^{i\theta} \right)
\end{equation}
where
\begin{equation}
H = qe^{i\gamma} + Fe^{i\Omega t}.
\end{equation}

The ansatz consists of expanding $f$ in Fourier series and  choosing the coefficients in terms of a single complex parameter $\nu(\omega,t)$ as:
\begin{equation}
	\label{eq::density}
	f(\omega,\theta,t) = \dfrac{g(\omega)}{2\pi} \left[ 1 + \sum_{n=1}^{\infty}\nu^n e^{-i n \theta} + c.c. \right ]
\end{equation} 
where $g(\omega)$ is the distribution of natural frequencies.

Defining $z = r e^{i\psi}$ and taking the continuum limit, Eq.(\ref{eq::orderP}) can be written as
\begin{equation}
	\label{eq::zspolo}
	z = \int f(\omega,\theta,t) e^{i\theta}d\theta d\omega = \int g(\omega) \nu(\omega) d\omega.
\end{equation}
This equation can be solved analytically for Lorentzian distributions $g(\omega) = \frac{1}{\pi}\frac{\Delta}{(\omega - \omega_{0})^2 + \Delta^2}$. In this case, the integrand has poles at $\omega = \omega_{0} \pm i\Delta$ and the overall integral can be performed by using the residues theorem, resulting in  $z = \nu(\omega_{0}+i\Delta)$ \cite{Ott2008}.

Using Eqs.\eqref{eq::velocity}-\eqref{eq::density}, we can now calculate the different terms in the continuity equation (\ref{eq::cont}):
\begin{align*}
	\dfrac{\partial f}{\partial t} &= \dfrac{g(\omega)}{2\pi} \left[ \sum_{n=1}^{\infty} n \dot{\nu} \nu^{n-1}e^{-in\theta} + c.c. \right], \\
	\dfrac{\partial f}{\partial \theta} &= -\dfrac{g(\omega)}{2\pi} \left[ \sum_{n=1}^{\infty} in  \nu^{n}e^{-in\theta} + c.c. \right], \\
	\dfrac{\partial v_{ \theta }}{\partial \theta} &= -\dfrac{1}{2} \left( He^{-i\theta} + H^*e^{i\theta} \right).
\end{align*}

Substituting in Eq. \eqref{eq::cont} and equating the coefficients of terms proportional to $e^{i n \theta}$ we consistently obtain the differential equation for the ansatz parameter $\nu$ 
\begin{equation}
	\label{eq::dyneta}
	\dot{\nu} =  i\omega \nu  + \dfrac{H}{2} - \dfrac{H^*}{2}\nu^2.
\end{equation}
Calculating all quantities at $\omega = \omega_{0}+i\Delta$ we can replace $\nu$  by $z$ in Eq. \eqref{eq::dyneta}. Also, 
using $\vec{q} \equiv \kk{} \vec{r}$ and Eq. \eqref{matrizK}, we can write $H$ as
\begin{equation}
	\label{eq::H}
H = K ze^{-i\alpha} - J z^*e^{-i\beta} + F e^{i\Omega t}
\end{equation}
resulting in 
\begin{equation}
	\label{eq:zcomplex2}
	\dot{z} = i(\omega_{0} + i\Delta)z - \dfrac{1}{2} \left [ \left( K z^*e^{i\alpha} -  Jze^{i\beta}  +F e^{-i\Omega t} \right) z^2  + \left(  K ze^{-i\alpha} - Jz^*e^{-i\beta} +F e^{i\Omega t} \right) \right] .
\end{equation}

Finally, separating real and imaginary parts of Eq. \eqref{eq:zcomplex2}, we find the dynamical equations for the module and phase of the order parameter $z$ as
\begin{equation}
		\dot{r} = -\Delta r + \dfrac{(1-r^2)}{2} \left[r K\cos{\alpha} - r J \cos{(2\psi + \beta)} + F \cos{(\psi-\Omega t)} \right]
\end{equation}
\begin{equation}
		r\dot{\psi} = +\omega_0 r - \dfrac{(1+r^2)}{2} \left[rK\sin{\alpha} - rJ \sin{(2\psi + \beta)} + F \sin{(\psi-\Omega t)} \right]
\end{equation}

Setting $\xi = \omega_0 - \Omega$ as the frequency mismatch between the oscillators and the external drive and defining $\phi = \psi - \Omega t$ we shift the time dependence to terms proportional to $J$:
\begin{equation}
	\dot{r} = -\Delta r + \dfrac{(1-r^2)}{2} \left[rK\cos{\alpha} - rJ \cos{(2\phi+ 2\Omega t+ \beta)} + F \cos{\phi} \right]
\end{equation}
\begin{equation}
	\dot{\phi} = \xi - \dfrac{(1+r^2)}{2} \left[K\sin{\alpha} - J \sin{(2\phi+ 2\Omega t+ \beta)} + \dfrac{F}{r} \sin{(\phi)}. \right]
\end{equation}

The non-linearity of the equations, combined with the explicit time dependence, makes analytical solutions very difficult to obtain. Thus, we shall resort to numerical analysis to further investigate this dynamics. We choose two sets of parameters, representing the two novel behaviors that the matrix coupled Kuramoto model present: the oscillatory and the phase tuned states.

\section{RESONANCES}

When investigating the dynamics of forced oscillators, we are interested in characterizing the possible synchronized states that can arise from their dynamics. Usually, there are two possible outcomes: the oscillators are in mutual synchronization, i.e., they synchronize spontaneously within their own dynamics, or forced synchronization, in which the oscillators adjust their frequencies with that of the periodic force to an integer ratio $n:m$. For the original Kuramoto model with external forcing the most common type of entrainment is $1:1$. However, since our dynamics have two explicit time scales, it is expected that more complex forms of synchronization can occur.

Formally, let us consider an oscillator under the action of a $T$-periodic external drive $F$
\begin{equation}
\dot{\theta} = g(\theta) + F(t), 
\end{equation}
in which $g(\theta + 2\pi) = g(\theta) \; \forall \; \theta$ and $F(t+T) = F(t) \; \forall \; t$. Thus, any $Q-$periodic motion of the oscillator, $\theta(t+Q) = \theta(t)$, will have a period that is related to $T$ by $Q = m T$. Defining $n$ as the number of times that the oscillator crosses $\theta = 0$ in one period, we say that the oscillators are $n:m$ synchronized to the external driving $F$.

In terms of bifurcation theory, the existence and stability of these synchronized states can be inferred if the function 
\begin{equation}
    h_{nm}(\theta) \equiv  h_m(\theta) - 2 \pi n - \theta,
\end{equation}
has a stable root, where $h_m$ is the $m-$ composition of the return map $h$. By changing the model parameters and recalculating $h_m(\theta)$, the function $h_{nm}$ undergoes a saddle-node bifurcation that indicates the boundaries of the $n:m$ synchronized region. However, analytical calculations of $h_{nm}$ are unfeasible in most cases, and we need to resort to numerical frameworks.

In this work, we will make use of the widely known synchronization index 
\begin{equation}
S_{nm} = \dfrac{1}{2\pi m (L-n)}\left|  \sum_{i=n+1}^{L} \phi_i - \phi_{i-n} - 2\pi n \right|
\end{equation}
to characterize possible $n:m$ synchronized states. Here, $L$ is the total number of crossings at $\theta = 0$ that occurs during the dynamics and $\phi_i$ is the phase of the external drive at the $i^{th}$ crossing. If the system is $n:m$ synchronized, them $S_{nm} \approx 0$. In this work we define this threshold as $S_{nm} < 10^{-5}$.

Another common way of characterizing periodic dynamics is using the winding or rotation number, defined as:
\begin{equation}
\mathcal{W} = \lim_{n \to \infty} \dfrac{\theta_n}{2\pi n},
\end{equation}
in which $\theta_n$ is the phase of the system on the $n$-th cycle of the driving force.

\section{DYNAMICAL BEHAVIOR AND ARNOLD TONGUES}

\subsection{OSCILLATORY STATES}

\begin{figure}[ht]
    \centering
    \includegraphics[width=0.95\textwidth]{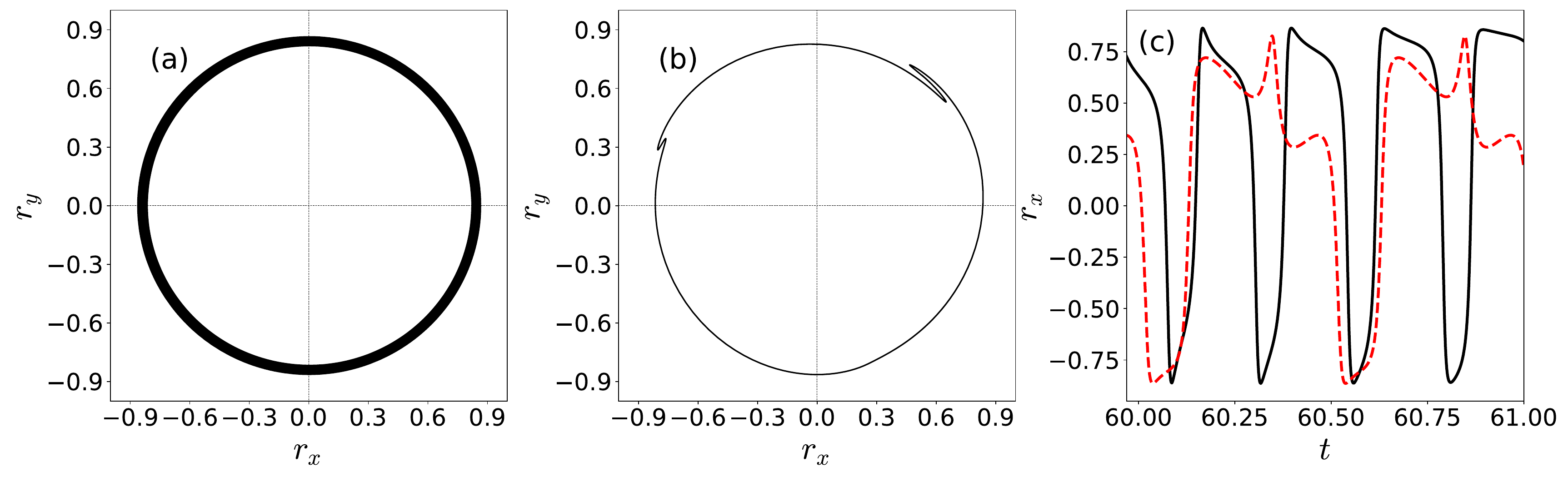}
    \caption{Examples of trajectories for the matrix coupled Kuramoto model with external forces with $F = 0.1$ and (a) $\xi = 0.01$ and (b) $\xi = 0.15$. Panel (c) shows the x component of $\vec{r}$ for cases (a) (black) and (b) (red) in a short time interval.}
    \label{fig:oscExample}
\end{figure}

As previously mentioned, for a given value of $\omega_0$, it is always possible to find a set of coupling matrices $\kk{}$ that drives the system to a specific synchronized state. Therefore, without loss of generality, we can fix $\omega_0=1$ and find coupling parameters that set the system in the oscillatory synchronized state, in which both the module and phase of the order parameter $\vec{r}$ oscillate in time. Specifically, we choose $K = 7.0$ ; $J = 1.0$ ; $\alpha = \beta = 0.0 $ and study the effect of external drive parameters $F$ and $\xi$ on the dynamics.  We found that $\vec{r}\,(t)$ displays two distinct types of behavior: (i) aperiodic trajectories, which span a certain area of the $F - \xi$ plane and, (ii) periodic orbits, as exemplified in Figure \ref{fig:oscExample}-(a) and (b). A further inspection of $x(t)$, Figure \ref{fig:oscExample}-(c), confirm these features. 

\begin{figure}[ht]
    \centering
    \includegraphics[width=0.75\textwidth]{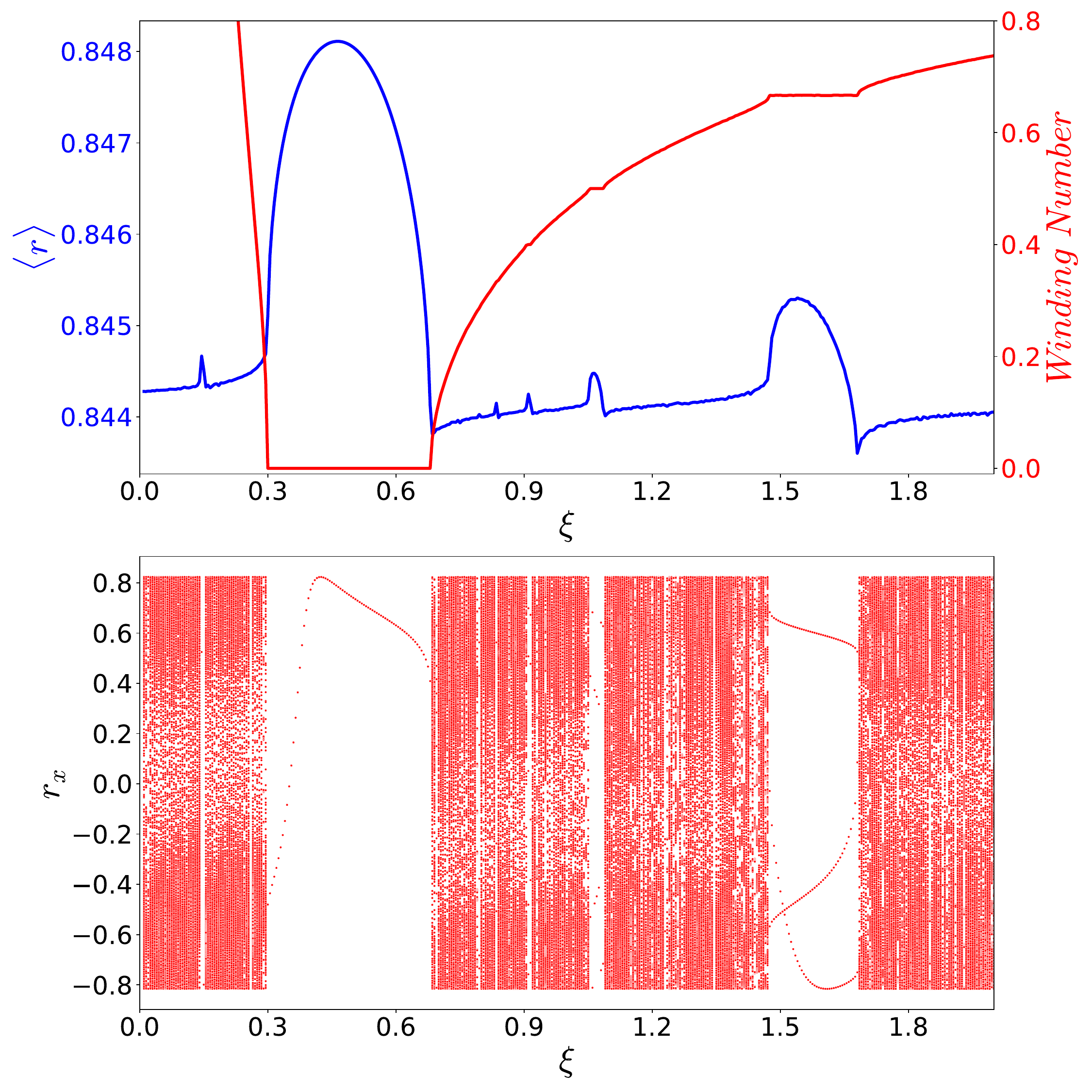}
    \caption{(a)Time-averaged order parameter $\langle r \rangle$ (left axis in blue) and Winding number (right axis in red) for different values of frequency drift $\xi$  and $F = 0.1$. (b) Values of the $x$-component of $\vec{r}$ found in the recurrence maps.}
    \label{fig:moundOscila}
\end{figure}

In order to characterize the behavior of the system across several values of $F$ and $\xi$, we first fixed $F = 0.1$ and changed  $\xi$ in the interval $[0,2]$. Figure \ref{fig:moundOscila}-(a) shows in blue (left y-axis) the time-averaged order parameter $\av{r}$. It can be seen that $\av{r}$ presents spikes and mounds for some intervals of $\xi$, indicating that qualitative changes in the dynamics might take place in the boundary of these regions. This suspicion is further confirmed by the value of the winding number $\mathcal{W}$, plotted in red (right y-axis). It is clear that the peaks in $\av{r}$ align perfectly with the regions  of constant $\mathcal{W}$. Visual inspection of the trajectories confirm that the plateaus in the winding number correspond to the periodic orbits. It is important to notice that the structure of the graph $\mathcal{W} \times \xi$ resembles closely the devil's staircases, which hints at the presence of Arnold tongues in the system.

To further understand the nature of these trajectories, we constructed the recurrence map, marking the value of $\vec{r}$ each time the external drive completes a cycle. From these maps, we extracted all values of $r_x$ for a fixed value of $\xi$, collecting them in Figure \ref{fig:moundOscila}-(b). This confirms that the plateaus of constant winding number consist of a small number of recurrence values, and thus to periodic orbits. Regions filled with a continuum of values, on the other hand, correspond to non-closed (or very high period) orbits. 

\begin{figure}[ht]
    \centering
    \includegraphics[width=0.5\textwidth]{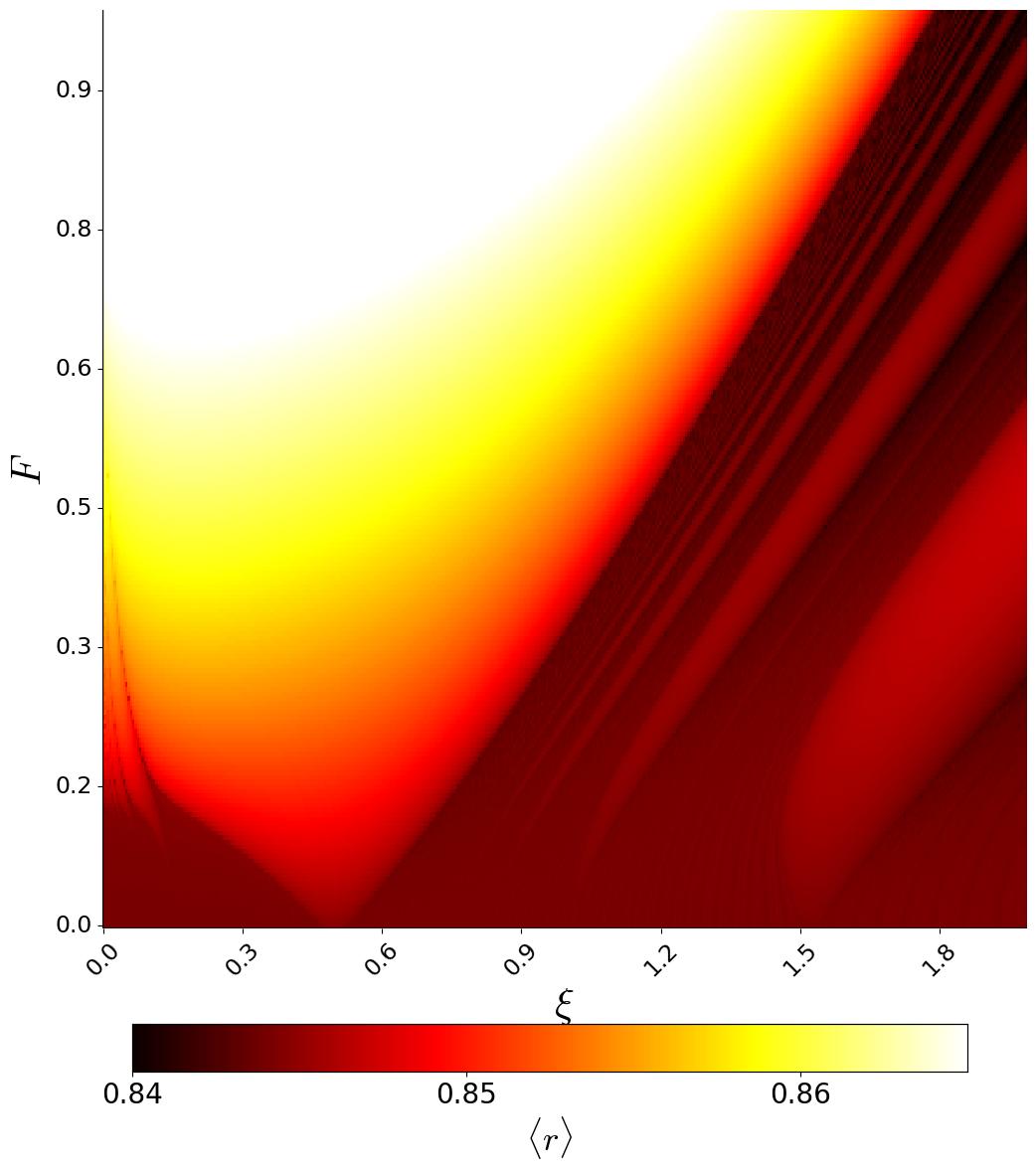}
    \caption{Heatmap of the time averaged order parameter $\av{r}$ for the oscillatory case.}
    \label{fig:heatk7j1Large}
\end{figure}

The next step is to see how these regions behave as we change the amplitude of the external force $F$. Figure \ref{fig:heatk7j1Large} shows a heatmap of $\av{r}$, varying both $F$ and $\xi$. It can be seen that, despite the changes in the time-averaged order parameter being small, a fringed pattern emerges. By inspecting the limit in which $F \to 0$, we see that the start of the larger stripes lies approximately on $\xi = 0.5$, $1.0$ and $1.5$.

\begin{figure}[ht]
    \centering
    \includegraphics[width=0.45\textwidth]{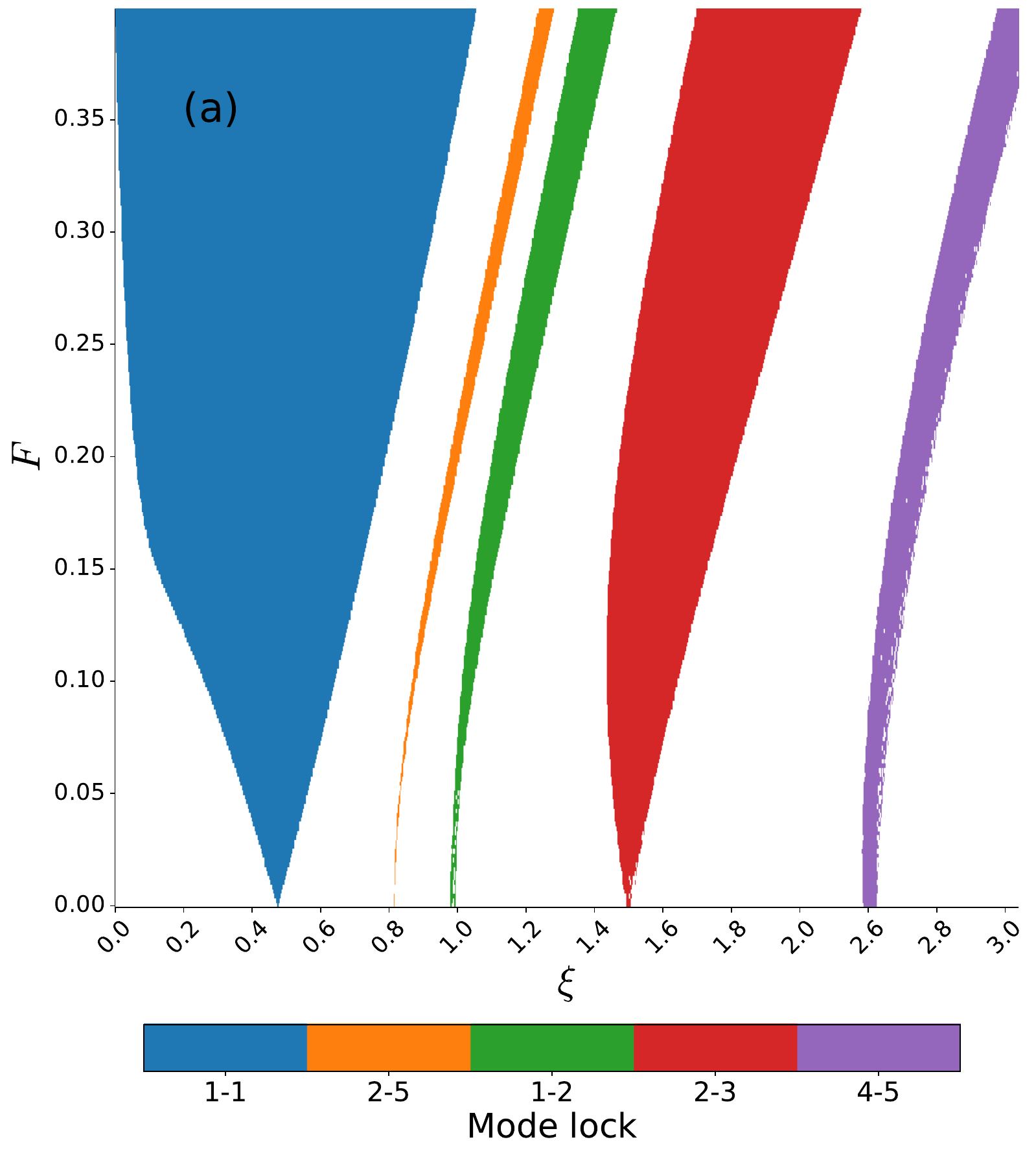}
    \includegraphics[width=0.45\textwidth]{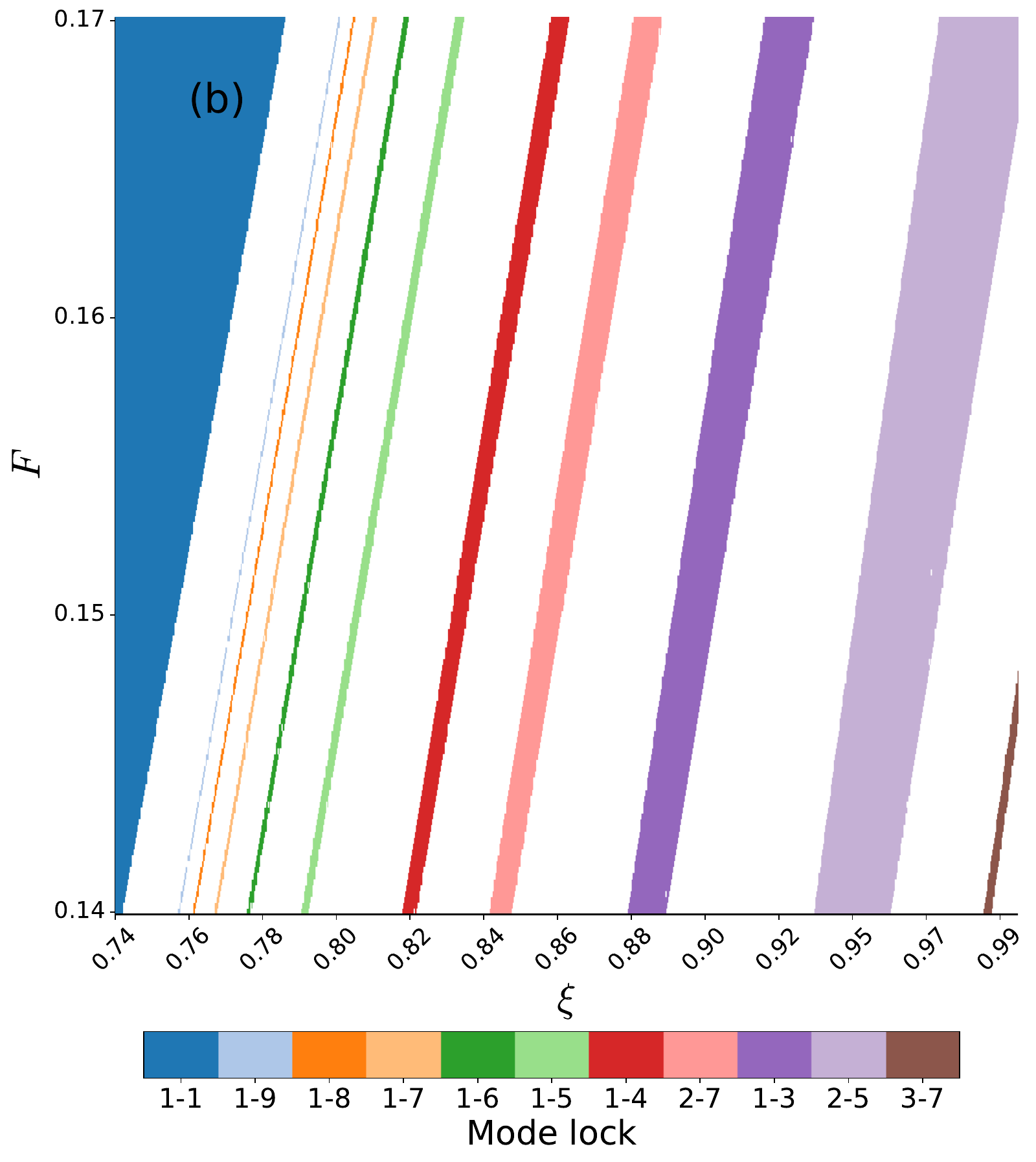}
    \caption{(a) Arnold tongues representing different mode locks in the matrix coupled Kuramoto model in the oscillatory state. (b) Zoom in an region of (a) showing a large variety of smaller tongues with different mode locks. (Color online)}
    \label{fig:modesk7j1}
\end{figure}

This fringed pattern hints at the existence of different types of synchronized states belonging to each fringe. Therefore, we calculated $S_{nm}$ for all integers in range $(1,10)$ and plotted each pair $n:m$ in different colors in Figure \ref{fig:modesk7j1}. Panel (a) shows the whole range of $\xi$, observing four major mode locks; the larger is $1:1$, as expected, but we also found minor stripes corresponding to modes $2:5$, $1:2$ and $2:3$. On the right panel, Figure \ref{fig:modesk7j1}-(b), we restrict ourselves to a smaller portion of the space, $0.74 < \xi < 1.00$ and found a myriad of smaller stripes corresponding to other $n:m$ synchronized states.

Although we have shown only the tongues for a single value of $K$, $J$ and $\omega$, similar patterns can be seen for all values of the coupling parameters that set the unforced system into the oscillatory states (not shown here). The major changes being the relative sizes of the peaks, since $\av{r}$ gets smaller as $K$ decreases,

\subsection{PHASE TUNED STATES}

In the previous subsection we studied the appearance of Arnold tongues in the oscillatory regime, where the external drive resonated with the internal rotation of the system. We now turn our attention to the phase tuned states, where the system stays locked to a given phase even for $\omega_0 \neq  0$. Here it is  not so evident that resonances can appear when the external drive parameters $F$ and $\Omega$ are varied. 

\begin{figure}[ht]
    \centering
    \includegraphics[width=0.70\textwidth]{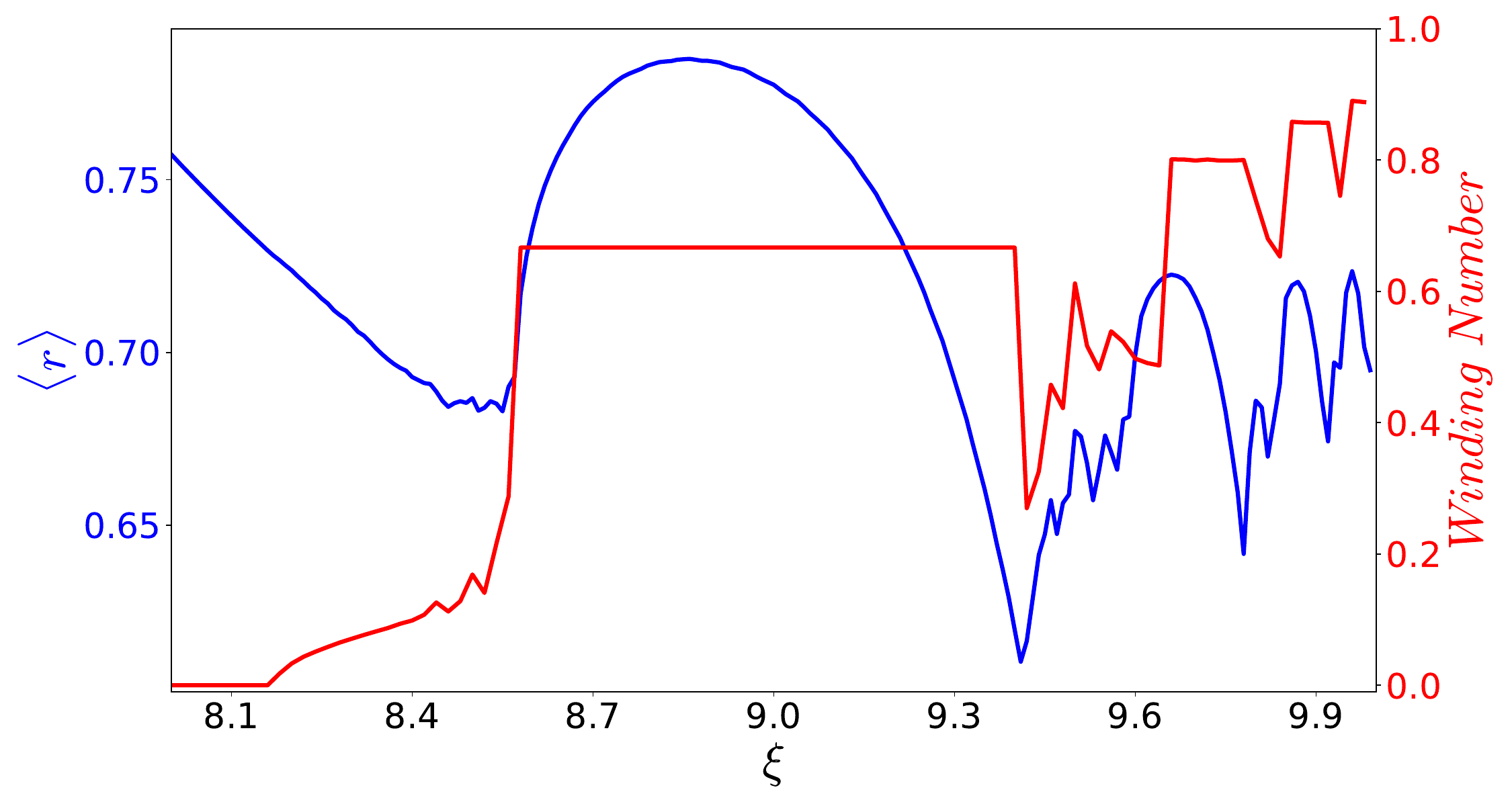}
    \includegraphics[width=0.65\textwidth]{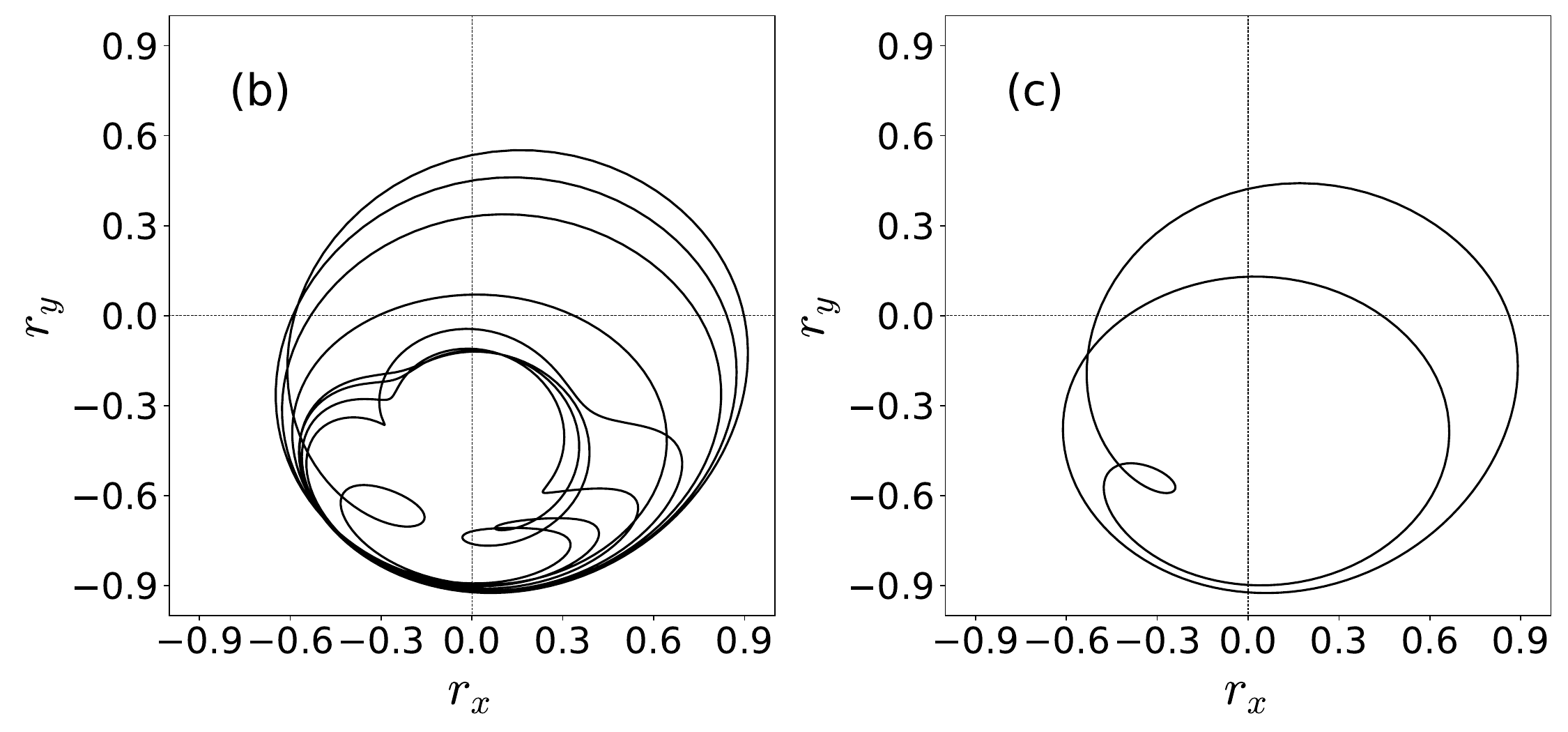}
    \caption{(a)Time-averaged order parameter $\langle r \rangle$ (left axis in blue) and Winding number (right axis in red) for different values of frequency drift $\xi$  and $F = 9.0$ (b-c) Examples of trajectories for the matrix coupled Kuramoto model set in the phase tuned states, with external forces with $F = 9.0$ and (b) $\xi = 9.54$ and (c) $\xi = 9.73$.}
    \label{fig:moundTuned}
\end{figure}

For this study we set the coupling parameters as $K = 7.0$ ; $J = 7.0$ ; $\alpha = \beta = 0.0 $, which set the system in a phase tuned state in the absence of external forces. Figure \ref{fig:moundTuned} shows the time averaged order parameter $\av{r}$ and the winding number $\mathcal{W}$ as a function of $\xi = \omega_0-\Omega$ for fixed $F = 9.0$ and $\omega_0 = 1.0$. It can be seen that the hill-like structure of $\av{r}$ is similar to those of Figure \ref{fig:moundOscila}, although less pronounced. The structure of the winding number $\mathcal{W}$ also shows plateaus separated by valleys, a distinct feature regarding the devil's  staircase found in the oscillatory case. It is important to notice that the alignment between the mounds in $\av{r}$ and the plateaus in $\mathcal{W}$ is not perfect for this case, indicating some underlying complication. By inspecting the trajectories of $\vec{r}$, Figs \ref{fig:moundTuned}-(b) and (c), it can be seen that the phase tuned direction, induced by the larger $J$ restricts some of the oscillations to a smaller portion of the phase space, contrasting with the trajectories of Fig. \ref{fig:oscExample}, that spans the four quadrants. Thus, the calculation of $\mathcal{W}$ may induce misleading results if the trajectories does not circulate around the origin.

\begin{figure}[ht]
    \centering
    \includegraphics[width=0.4\textwidth]{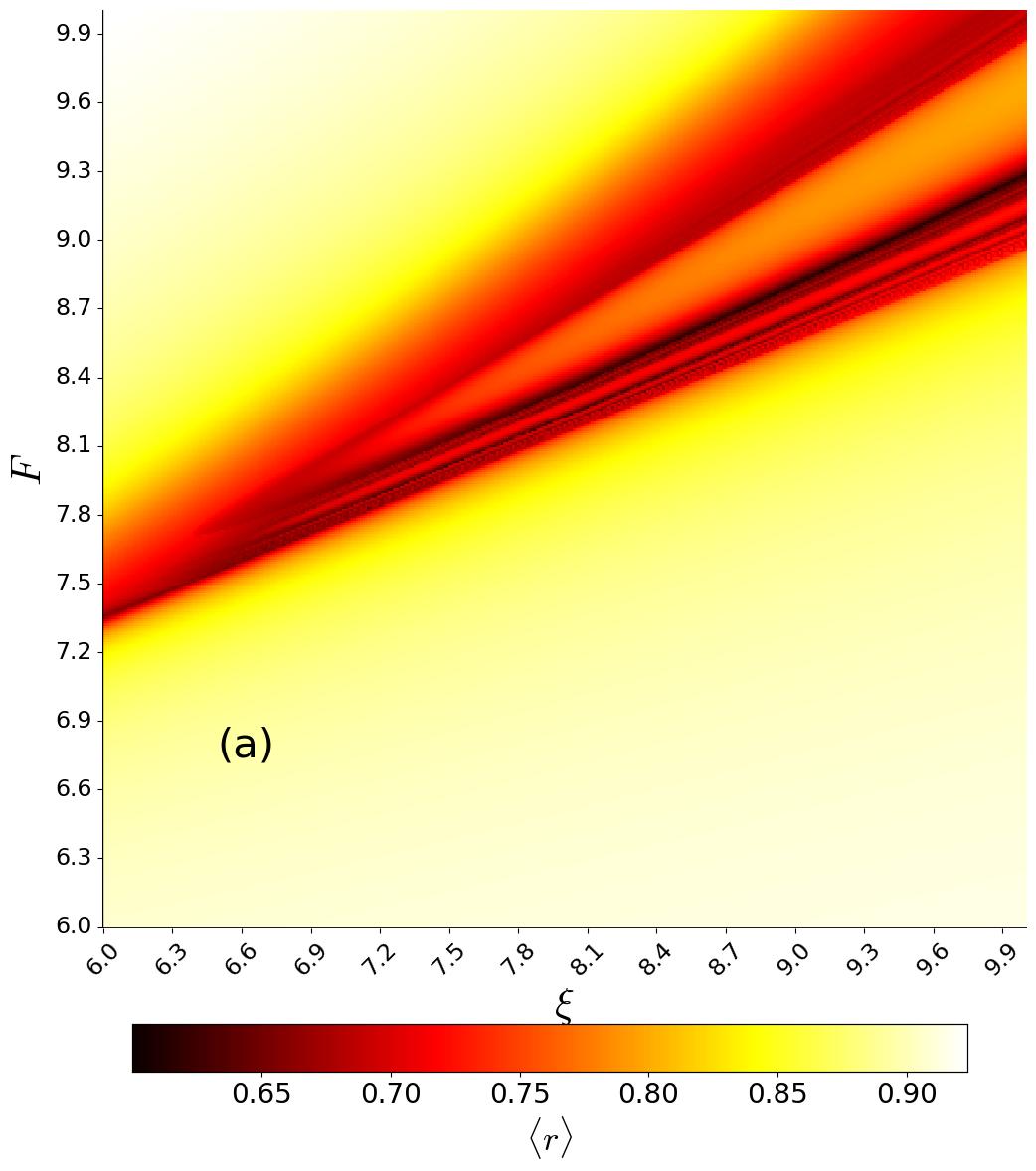}
    \includegraphics[width=0.422\textwidth]{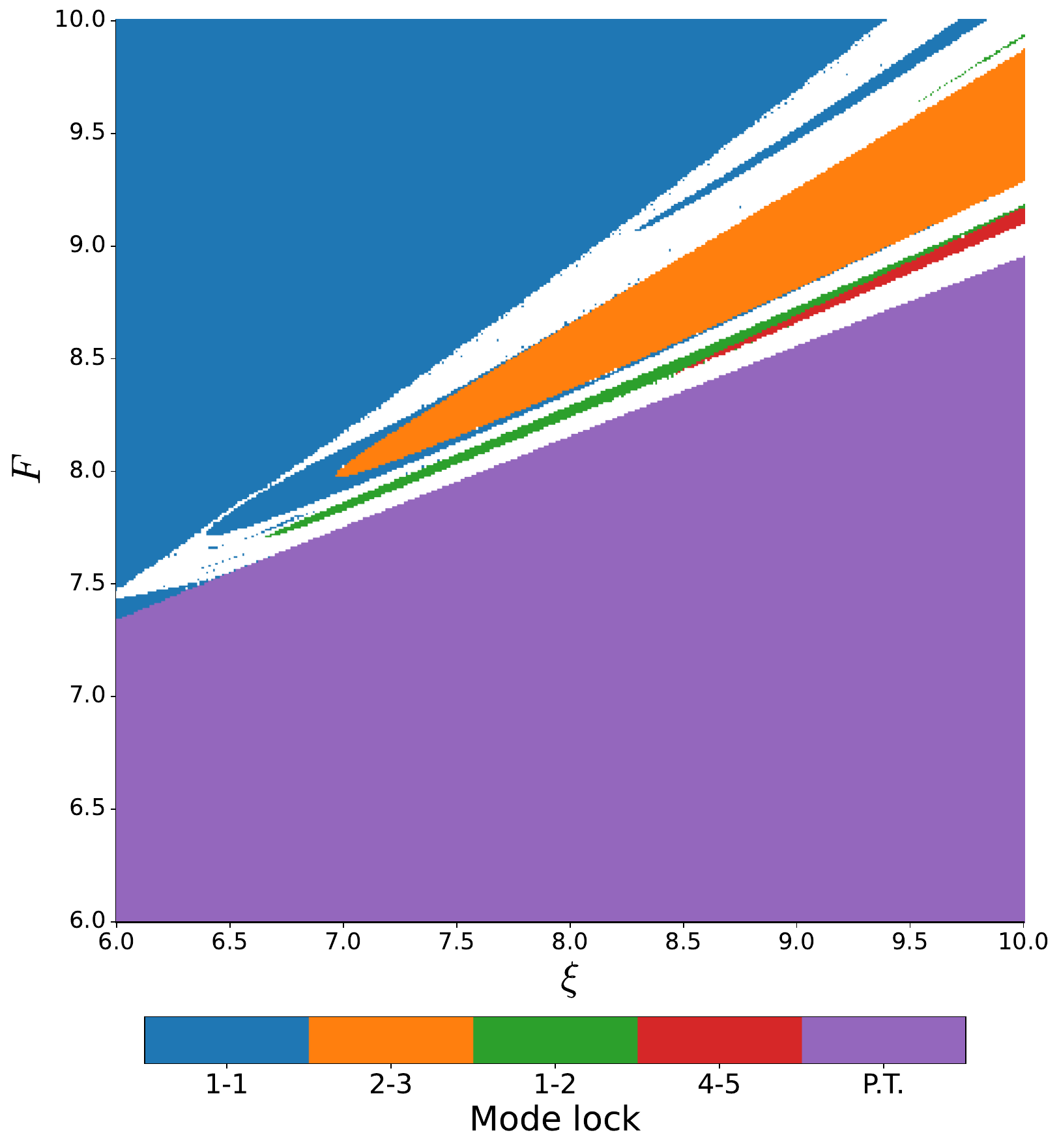}
    \caption{(a) Heatmap of the time averaged order parameter $\av{r}$ for the phase tuned case. (b) Arnold tongues representing different mode locks in the matrix coupled Kuramoto model in the phase tuned (P.T.) state.}
    \label{fig:heatk7j7Large}
\end{figure}

We also constructed the $\av{r}$ heatmap for the phase tuned case, identifying possible entrainment modes by using the synchronization index $S_{nm}$. Although similar fringed patterns can be seen in Figs \ref{fig:heatk7j7Large}, there are some notable distinctions between them and those found in the oscillatory case. First, some tongues are weirdly shaped, such as the 1:2 mode lock (green region in the Figure), that arises out of the thin air and collapses on the 4:5 mode. In addition, some modes appears in more than one tongue, such as the thin green stripe on the top right corner of Fig. \ref{fig:heatk7j7Large}-(b).  Therefore, despite the entrainment modes being similar to the oscillatory case, the diagram and relations among them are quite different, with some of the tongues constrained to very small parameter regions.

\section{FINAL REMARKS}

In this work we investigated the matrix coupled Kuramoto model under the influences of external periodic forces. We provided a full description of the system in the limit of infinitely many oscillators. Using of the Ott-Antonsen ansatz we performed the corresponding dimensional reduction and found the dynamical equations for the order parameter. Since the equations are time-dependent and cannot be solved analytically, we performed extensive numerical calculations and uncovered multiple synchronization modes. In addition of the common $1:1$ mode, in which the oscillators synchronize their dynamics exactly to the external drive, we found extensive regions of several $n:m$ modes that give birth to multiple Arnold tongues, regions of the phase space that present constant rotation of winding number, despite changes in the frequencies of the external drives.

In order to understand why the different entrainment modes arise in the matrix coupled model, it is useful to rewrite the dynamical equation in vector form, Eq.(\ref{eq::kuraFrustMF}), separating the symmetric and anti-symmetric components of $\kk{}$:
\begin{equation}
	\dfrac{d \sig{i}{}}{dt} = \left[ \textbf{W}_i \sig{i}{} + \kk{R}\,\vec{r} - (\sig{i}{}\cdot \kk{R}\,\vec{r}) \sig{i}{}\right]  + \left[ \kk{S}\,\vec{r} - (\sig{i}{}\cdot \kk{S}\,\vec{r}) \sig{i}{} \right]
\end{equation}

in which the component $\kk{R}$ corresponds to the usual Kuramoto-Sakaguchi model \cite{Sakaguchi1988} and $\kk{S}$ is the new interaction, parametrized by $J$ and $\beta$ (see Eqs. (\ref{matrizK})-(\ref{eq::thetai})). Comparing with Eq.(\ref{eq::kuraFrustForced}), we see that $\kk{S}\, \vec{r}$ can be interpreted as an internal time-independent force that breaks the rotational symmetry of the system, since its value depends on the direction of $\vec{r}$. This is similar to the perturbed Kicked Rotor, where the intensity of the kicks depend on the state of the rotor \cite{escande1985stochasticity}, although here the force acts continuously.

When an external periodic force is added to the equation, the resulting system can be seen as a set of Kuramoto-Sakaguchi oscillators under the influence of two forces, one internal and aperiodic, represented by $J$ and $\beta$, and the other, external and periodic, modulated by $F$ and $\Omega$. The simultaneous action of these two forces makes it impossible to find a frame of reference that rotates simultaneously with both, since they usually have distinct frequencies. In addition, the internal force is not periodic. Arnold tongues arise when the frequencies of these forces are rationally connected. In the original Kuramoto it would take  two external periodic forces of distinct frequencies to produce similar effects. 

These novel phenomena of mode synchronization, that do not occur on the original Kuramoto model, can be seen on some biological oscilators \cite{Jimenez2022} and the segmentation clock of embryos \cite{Sanchez2022}, highlighting the importance of investigating the matrix coupled extension of the Kuramoto model. A natural prospect of this study is to consider different substrates beyond the all-to-all interactions \cite{Rodrigues2016,odor2023}.

\begin{acknowledgements}
 This work was partly supported by FAPESP [Grant 2023/03917-4 (G.S.C.)] and [Grant 2021/ 14335-0 (M.A.M.A.)] and by CNPq, Brazil, [Grant 303814/2023-3 (M.A.M.A.)].
\end{acknowledgements}


\begin{thebibliography}{47}%
	\makeatletter
	\providecommand \@ifxundefined [1]{%
		\@ifx{#1\undefined}
	}%
	\providecommand \@ifnum [1]{%
		\ifnum #1\expandafter \@firstoftwo
		\else \expandafter \@secondoftwo
		\fi
	}%
	\providecommand \@ifx [1]{%
		\ifx #1\expandafter \@firstoftwo
		\else \expandafter \@secondoftwo
		\fi
	}%
	\providecommand \natexlab [1]{#1}%
	\providecommand \enquote  [1]{``#1''}%
	\providecommand \bibnamefont  [1]{#1}%
	\providecommand \bibfnamefont [1]{#1}%
	\providecommand \citenamefont [1]{#1}%
	\providecommand \href@noop [0]{\@secondoftwo}%
	\providecommand \href [0]{\begingroup \@sanitize@url \@href}%
	\providecommand \@href[1]{\@@startlink{#1}\@@href}%
	\providecommand \@@href[1]{\endgroup#1\@@endlink}%
	\providecommand \@sanitize@url [0]{\catcode `\\12\catcode `\$12\catcode
		`\&12\catcode `\#12\catcode `\^12\catcode `\_12\catcode `\%12\relax}%
	\providecommand \@@startlink[1]{}%
	\providecommand \@@endlink[0]{}%
	\providecommand \url  [0]{\begingroup\@sanitize@url \@url }%
	\providecommand \@url [1]{\endgroup\@href {#1}{\urlprefix }}%
	\providecommand \urlprefix  [0]{URL }%
	\providecommand \Eprint [0]{\href }%
	\providecommand \doibase [0]{https://doi.org/}%
	\providecommand \selectlanguage [0]{\@gobble}%
	\providecommand \bibinfo  [0]{\@secondoftwo}%
	\providecommand \bibfield  [0]{\@secondoftwo}%
	\providecommand \translation [1]{[#1]}%
	\providecommand \BibitemOpen [0]{}%
	\providecommand \bibitemStop [0]{}%
	\providecommand \bibitemNoStop [0]{.\EOS\space}%
	\providecommand \EOS [0]{\spacefactor3000\relax}%
	\providecommand \BibitemShut  [1]{\csname bibitem#1\endcsname}%
	\let\auto@bib@innerbib\@empty
	\bibitem [{\citenamefont {Pikovsky}\ \emph {et~al.}(2001)\citenamefont
		{Pikovsky}, \citenamefont {Rosenblum},\ and\ \citenamefont
		{Kurths}}]{pikovsky2001synchronization}%
	\BibitemOpen
	\bibfield  {author} {\bibinfo {author} {\bibfnamefont {A.}~\bibnamefont
			{Pikovsky}}, \bibinfo {author} {\bibfnamefont {M.}~\bibnamefont
			{Rosenblum}},\ and\ \bibinfo {author} {\bibfnamefont {J.}~\bibnamefont
			{Kurths}},\ }\bibfield  {title} {\bibinfo {title} {Synchronization: A
			universal concept in nonlinear sciences},\ }\href@noop {} {\bibfield
		{journal} {\bibinfo  {journal} {Cambridge University Press}\ }\textbf
		{\bibinfo {volume} {12}} (\bibinfo {year} {2001})}\BibitemShut {NoStop}%
	\bibitem [{\citenamefont {Boccaletti}\ \emph {et~al.}(2018)\citenamefont
		{Boccaletti}, \citenamefont {Pisarchik}, \citenamefont {Del~Genio},\ and\
		\citenamefont {Amann}}]{boccaletti2018synchronization}%
	\BibitemOpen
	\bibfield  {author} {\bibinfo {author} {\bibfnamefont {S.}~\bibnamefont
			{Boccaletti}}, \bibinfo {author} {\bibfnamefont {A.~N.}\ \bibnamefont
			{Pisarchik}}, \bibinfo {author} {\bibfnamefont {C.~I.}\ \bibnamefont
			{Del~Genio}},\ and\ \bibinfo {author} {\bibfnamefont {A.}~\bibnamefont
			{Amann}},\ }\href@noop {} {\emph {\bibinfo {title} {Synchronization: from
				coupled systems to complex networks}}}\ (\bibinfo  {publisher} {Cambridge
		University Press},\ \bibinfo {year} {2018})\BibitemShut {NoStop}%
	\bibitem [{\citenamefont {Michaels}\ \emph {et~al.}(1987)\citenamefont
		{Michaels}, \citenamefont {Matyas},\ and\ \citenamefont
		{Jalife}}]{Michaels704}%
	\BibitemOpen
	\bibfield  {author} {\bibinfo {author} {\bibfnamefont {D.~C.}\ \bibnamefont
			{Michaels}}, \bibinfo {author} {\bibfnamefont {E.~P.}\ \bibnamefont
			{Matyas}},\ and\ \bibinfo {author} {\bibfnamefont {J.}~\bibnamefont
			{Jalife}},\ }\bibfield  {title} {\bibinfo {title} {Mechanisms of sinoatrial
			pacemaker synchronization: a new hypothesis.},\ }\href
	{https://doi.org/10.1161/01.RES.61.5.704} {\bibfield  {journal} {\bibinfo
			{journal} {Circulation Research}\ }\textbf {\bibinfo {volume} {61}},\
		\bibinfo {pages} {704} (\bibinfo {year} {1987})},\ \Eprint
	{https://arxiv.org/abs/http://circres.ahajournals.org/content/61/5/704.full.pdf}
	{http://circres.ahajournals.org/content/61/5/704.full.pdf} \BibitemShut
	{NoStop}%
	\bibitem [{\citenamefont {Wu}\ \emph {et~al.}(2023)\citenamefont {Wu},
		\citenamefont {Peng}, \citenamefont {Boscolo}, \citenamefont {Finot},\ and\
		\citenamefont {Zeng}}]{Xiuqi2023}%
	\BibitemOpen
	\bibfield  {author} {\bibinfo {author} {\bibfnamefont {X.}~\bibnamefont
			{Wu}}, \bibinfo {author} {\bibfnamefont {J.}~\bibnamefont {Peng}}, \bibinfo
		{author} {\bibfnamefont {S.}~\bibnamefont {Boscolo}}, \bibinfo {author}
		{\bibfnamefont {C.}~\bibnamefont {Finot}},\ and\ \bibinfo {author}
		{\bibfnamefont {H.}~\bibnamefont {Zeng}},\ }\bibfield  {title} {\bibinfo
		{title} {Synchronization, desynchronization, and intermediate regime of
			breathing solitons and soliton molecules in a laser cavity},\ }\href
	{https://doi.org/10.1103/PhysRevLett.131.263802} {\bibfield  {journal}
		{\bibinfo  {journal} {Phys. Rev. Lett.}\ }\textbf {\bibinfo {volume} {131}},\
		\bibinfo {pages} {263802} (\bibinfo {year} {2023})}\BibitemShut {NoStop}%
	\bibitem [{\citenamefont {Winfree}(1967)}]{Winfree1967}%
	\BibitemOpen
	\bibfield  {author} {\bibinfo {author} {\bibfnamefont {A.~T.}\ \bibnamefont
			{Winfree}},\ }\bibfield  {title} {\bibinfo {title} {Biological rhythms and
			the behavior of populations of coupled oscillators},\ }\href
	{https://doi.org/https://doi.org/10.1016/0022-5193(67)90051-3} {\bibfield
		{journal} {\bibinfo  {journal} {Journal of Theoretical Biology}\ }\textbf
		{\bibinfo {volume} {16}},\ \bibinfo {pages} {15} (\bibinfo {year}
		{1967})}\BibitemShut {NoStop}%
	\bibitem [{\citenamefont {Kuramoto}(1975)}]{kuramoto1975}%
	\BibitemOpen
	\bibfield  {author} {\bibinfo {author} {\bibfnamefont {Y.}~\bibnamefont
			{Kuramoto}},\ }\bibfield  {title} {\bibinfo {title} {Self-entrainment of a
			population of coupled non-linear oscillators},\ }in\ \href@noop {} {\emph
		{\bibinfo {booktitle} {International Symposium on Mathematical Problems in
				Theoretical Physics}}},\ \bibinfo {editor} {edited by\ \bibinfo {editor}
		{\bibfnamefont {H.}~\bibnamefont {Araki}}}\ (\bibinfo  {publisher} {Springer
		Berlin Heidelberg},\ \bibinfo {address} {Berlin, Heidelberg},\ \bibinfo
	{year} {1975})\ pp.\ \bibinfo {pages} {420--422}\BibitemShut {NoStop}%
	\bibitem [{\citenamefont {Skardal}\ and\ \citenamefont
		{Arenas}(2020)}]{Skardal2020}%
	\BibitemOpen
	\bibfield  {author} {\bibinfo {author} {\bibfnamefont {P.~S.}\ \bibnamefont
			{Skardal}}\ and\ \bibinfo {author} {\bibfnamefont {A.}~\bibnamefont
			{Arenas}},\ }\bibfield  {title} {\bibinfo {title} {Higher order interactions
			in complex networks of phase oscillators promote abrupt synchronization
			switching},\ }\href {https://doi.org/10.1038/s42005-020-00485-0} {\bibfield
		{journal} {\bibinfo  {journal} {Communications Physics}\ }\textbf {\bibinfo
			{volume} {3}},\ \bibinfo {pages} {1} (\bibinfo {year} {2020})}\BibitemShut
	{NoStop}%
	\bibitem [{\citenamefont {Chandrasekar}\ \emph {et~al.}(2020)\citenamefont
		{Chandrasekar}, \citenamefont {Manoranjani},\ and\ \citenamefont
		{Gupta}}]{Chandrasekar2020}%
	\BibitemOpen
	\bibfield  {author} {\bibinfo {author} {\bibfnamefont {V.~K.}\ \bibnamefont
			{Chandrasekar}}, \bibinfo {author} {\bibfnamefont {M.}~\bibnamefont
			{Manoranjani}},\ and\ \bibinfo {author} {\bibfnamefont {S.}~\bibnamefont
			{Gupta}},\ }\bibfield  {title} {\bibinfo {title} {Kuramoto model in the
			presence of additional interactions that break rotational symmetry},\ }\href
	{https://doi.org/10.1103/PhysRevE.102.012206} {\bibfield  {journal} {\bibinfo
			{journal} {Phys. Rev. E}\ }\textbf {\bibinfo {volume} {102}},\ \bibinfo
		{pages} {012206} (\bibinfo {year} {2020})}\BibitemShut {NoStop}%
	\bibitem [{\citenamefont {Liz\'arraga}\ and\ \citenamefont
		{de~Aguiar}(2023)}]{Lizarraga2023}%
	\BibitemOpen
	\bibfield  {author} {\bibinfo {author} {\bibfnamefont {J.~U.~F.}\
			\bibnamefont {Liz\'arraga}}\ and\ \bibinfo {author} {\bibfnamefont
			{M.~A.~M.}\ \bibnamefont {de~Aguiar}},\ }\bibfield  {title} {\bibinfo {title}
		{Synchronization of sakaguchi swarmalators},\ }\href
	{https://doi.org/10.1103/PhysRevE.108.024212} {\bibfield  {journal} {\bibinfo
			{journal} {Phys. Rev. E}\ }\textbf {\bibinfo {volume} {108}},\ \bibinfo
		{pages} {024212} (\bibinfo {year} {2023})}\BibitemShut {NoStop}%
	\bibitem [{\citenamefont {Costa}\ \emph {et~al.}(2025)\citenamefont {Costa},
		\citenamefont {Novaes},\ and\ \citenamefont {{de Aguiar}}}]{Costa2025}%
	\BibitemOpen
	\bibfield  {author} {\bibinfo {author} {\bibfnamefont {G.~S.}\ \bibnamefont
			{Costa}}, \bibinfo {author} {\bibfnamefont {M.}~\bibnamefont {Novaes}},\ and\
		\bibinfo {author} {\bibfnamefont {M.~A.}\ \bibnamefont {{de Aguiar}}},\
	}\bibfield  {title} {\bibinfo {title} {Exact solutions of the kuramoto model
			with asymmetric higher order interactions of arbitrary order},\ }\href
	{https://doi.org/https://doi.org/10.1016/j.chaos.2025.116243} {\bibfield
		{journal} {\bibinfo  {journal} {Chaos, Solitons \& Fractals}\ }\textbf
		{\bibinfo {volume} {195}},\ \bibinfo {pages} {116243} (\bibinfo {year}
		{2025})}\BibitemShut {NoStop}%
	\bibitem [{\citenamefont {Murayama}\ \emph {et~al.}(2017)\citenamefont
		{Murayama}, \citenamefont {Kori}, \citenamefont {Oshima}, \citenamefont
		{Kondo}, \citenamefont {Iwasaki},\ and\ \citenamefont {Ito}}]{Murayama2017}%
	\BibitemOpen
	\bibfield  {author} {\bibinfo {author} {\bibfnamefont {Y.}~\bibnamefont
			{Murayama}}, \bibinfo {author} {\bibfnamefont {H.}~\bibnamefont {Kori}},
		\bibinfo {author} {\bibfnamefont {C.}~\bibnamefont {Oshima}}, \bibinfo
		{author} {\bibfnamefont {T.}~\bibnamefont {Kondo}}, \bibinfo {author}
		{\bibfnamefont {H.}~\bibnamefont {Iwasaki}},\ and\ \bibinfo {author}
		{\bibfnamefont {H.}~\bibnamefont {Ito}},\ }\bibfield  {title} {\bibinfo
		{title} {Low temperature nullifies the circadian clock in cyanobacteria
			through hopf bifurcation},\ }\href {https://doi.org/10.1073/pnas.1620378114}
	{\bibfield  {journal} {\bibinfo  {journal} {Proceedings of the National
				Academy of Sciences}\ }\textbf {\bibinfo {volume} {114}},\ \bibinfo {pages}
		{5641} (\bibinfo {year} {2017})},\ \Eprint
	{https://arxiv.org/abs/https://www.pnas.org/doi/pdf/10.1073/pnas.1620378114}
	{https://www.pnas.org/doi/pdf/10.1073/pnas.1620378114} \BibitemShut {NoStop}%
	\bibitem [{\citenamefont {Prokkola}\ and\ \citenamefont
		{Nikinmaa}(2018)}]{Prokkola2018}%
	\BibitemOpen
	\bibfield  {author} {\bibinfo {author} {\bibfnamefont {J.~M.}\ \bibnamefont
			{Prokkola}}\ and\ \bibinfo {author} {\bibfnamefont {M.}~\bibnamefont
			{Nikinmaa}},\ }\bibfield  {title} {\bibinfo {title} {Circadian rhythms and
			environmental disturbances - underexplored interactions},\ }\href
	{https://doi.org/10.1242/jeb.179267} {\bibfield  {journal} {\bibinfo
			{journal} {Journal of Experimental Biology}\ }\textbf {\bibinfo {volume}
			{221}},\ \bibinfo {pages} {jeb179267} (\bibinfo {year} {2018})},\ \Eprint
	{https://arxiv.org/abs/https://journals.biologists.com/jeb/article-pdf/221/16/jeb179267/1903366/jeb179267.pdf}
	{https://journals.biologists.com/jeb/article-pdf/221/16/jeb179267/1903366/jeb179267.pdf}
	\BibitemShut {NoStop}%
	\bibitem [{\citenamefont {Jimenez}\ \emph {et~al.}(2022)\citenamefont
		{Jimenez}, \citenamefont {Lu}, \citenamefont {Jambhekar},\ and\ \citenamefont
		{Lahav}}]{Jimenez2022}%
	\BibitemOpen
	\bibfield  {author} {\bibinfo {author} {\bibfnamefont {A.}~\bibnamefont
			{Jimenez}}, \bibinfo {author} {\bibfnamefont {Y.}~\bibnamefont {Lu}},
		\bibinfo {author} {\bibfnamefont {A.}~\bibnamefont {Jambhekar}},\ and\
		\bibinfo {author} {\bibfnamefont {G.}~\bibnamefont {Lahav}},\ }\bibfield
	{title} {\bibinfo {title} {Principles, mechanisms and functions of
			entrainment in biological oscillators},\ }\href
	{https://doi.org/10.1098/rsfs.2021.0088} {\bibfield  {journal} {\bibinfo
			{journal} {Interface Focus}\ }\textbf {\bibinfo {volume} {12}},\ \bibinfo
		{pages} {20210088} (\bibinfo {year} {2022})},\ \Eprint
	{https://arxiv.org/abs/https://royalsocietypublishing.org/rsfs/article-pdf/doi/10.1098/rsfs.2021.0088/1468342/rsfs.2021.0088.pdf}
	{https://royalsocietypublishing.org/rsfs/article-pdf/doi/10.1098/rsfs.2021.0088/1468342/rsfs.2021.0088.pdf}
	\BibitemShut {NoStop}%
	\bibitem [{\citenamefont {Sanchez}\ \emph {et~al.}(2022)\citenamefont
		{Sanchez}, \citenamefont {Mochulska}, \citenamefont {Mauffette~Denis},
		\citenamefont {Mönke}, \citenamefont {Tomita}, \citenamefont
		{Tsuchida-Straeten}, \citenamefont {Petersen}, \citenamefont {Sonnen},
		\citenamefont {François},\ and\ \citenamefont {Aulehla}}]{Sanchez2022}%
	\BibitemOpen
	\bibfield  {author} {\bibinfo {author} {\bibfnamefont {P.~G.~L.}\
			\bibnamefont {Sanchez}}, \bibinfo {author} {\bibfnamefont {V.}~\bibnamefont
			{Mochulska}}, \bibinfo {author} {\bibfnamefont {C.}~\bibnamefont
			{Mauffette~Denis}}, \bibinfo {author} {\bibfnamefont {G.}~\bibnamefont
			{Mönke}}, \bibinfo {author} {\bibfnamefont {T.}~\bibnamefont {Tomita}},
		\bibinfo {author} {\bibfnamefont {N.}~\bibnamefont {Tsuchida-Straeten}},
		\bibinfo {author} {\bibfnamefont {Y.}~\bibnamefont {Petersen}}, \bibinfo
		{author} {\bibfnamefont {K.}~\bibnamefont {Sonnen}}, \bibinfo {author}
		{\bibfnamefont {P.}~\bibnamefont {François}},\ and\ \bibinfo {author}
		{\bibfnamefont {A.}~\bibnamefont {Aulehla}},\ }\bibfield  {title} {\bibinfo
		{title} {Arnold tongue entrainment reveals dynamical principles of the
			embryonic segmentation clock},\ }\href {https://doi.org/10.7554/eLife.79575}
	{\bibfield  {journal} {\bibinfo  {journal} {eLife}\ }\textbf {\bibinfo
			{volume} {11}},\ \bibinfo {pages} {e79575} (\bibinfo {year}
		{2022})}\BibitemShut {NoStop}%
	\bibitem [{\citenamefont {Shim}\ \emph {et~al.}(2007)\citenamefont {Shim},
		\citenamefont {Imboden},\ and\ \citenamefont {Mohanty}}]{Shim2007}%
	\BibitemOpen
	\bibfield  {author} {\bibinfo {author} {\bibfnamefont {S.-B.}\ \bibnamefont
			{Shim}}, \bibinfo {author} {\bibfnamefont {M.}~\bibnamefont {Imboden}},\ and\
		\bibinfo {author} {\bibfnamefont {P.}~\bibnamefont {Mohanty}},\ }\bibfield
	{title} {\bibinfo {title} {Synchronized oscillation in coupled nanomechanical
			oscillators},\ }\href {https://doi.org/10.1126/science.1137307} {\bibfield
		{journal} {\bibinfo  {journal} {Science}\ }\textbf {\bibinfo {volume}
			{316}},\ \bibinfo {pages} {95} (\bibinfo {year} {2007})},\ \Eprint
	{https://arxiv.org/abs/https://www.science.org/doi/pdf/10.1126/science.1137307}
	{https://www.science.org/doi/pdf/10.1126/science.1137307} \BibitemShut
	{NoStop}%
	\bibitem [{\citenamefont {Goldin}\ and\ \citenamefont
		{Kasimov}(2022)}]{Kasimov2022}%
	\BibitemOpen
	\bibfield  {author} {\bibinfo {author} {\bibfnamefont {A.~Y.}\ \bibnamefont
			{Goldin}}\ and\ \bibinfo {author} {\bibfnamefont {A.~R.}\ \bibnamefont
			{Kasimov}},\ }\bibfield  {title} {\bibinfo {title} {Synchronization of
			detonations: Arnold tongues and devil's staircases},\ }\href
	{https://doi.org/10.1017/jfm.2022.581} {\bibfield  {journal} {\bibinfo
			{journal} {Journal of Fluid Mechanics}\ }\textbf {\bibinfo {volume} {946}},\
		\bibinfo {pages} {R1} (\bibinfo {year} {2022})}\BibitemShut {NoStop}%
	\bibitem [{\citenamefont {Ali}\ \emph {et~al.}(2024)\citenamefont {Ali},
		\citenamefont {Chen},\ and\ \citenamefont {Radhakrishnan}}]{Ali2024}%
	\BibitemOpen
	\bibfield  {author} {\bibinfo {author} {\bibfnamefont {M.~M.}\ \bibnamefont
			{Ali}}, \bibinfo {author} {\bibfnamefont {P.-W.}\ \bibnamefont {Chen}},\ and\
		\bibinfo {author} {\bibfnamefont {C.}~\bibnamefont {Radhakrishnan}},\
	}\bibfield  {title} {\bibinfo {title} {Detecting quantum phase localization
			using arnold tongue},\ }\href
	{https://doi.org/https://doi.org/10.1016/j.physa.2023.129436} {\bibfield
		{journal} {\bibinfo  {journal} {Physica A: Statistical Mechanics and its
				Applications}\ }\textbf {\bibinfo {volume} {633}},\ \bibinfo {pages} {129436}
		(\bibinfo {year} {2024})}\BibitemShut {NoStop}%
	\bibitem [{\citenamefont {Sakaguchi}\ and\ \citenamefont
		{Kuramoto}(1986)}]{Sakaguchi1986}%
	\BibitemOpen
	\bibfield  {author} {\bibinfo {author} {\bibfnamefont {H.}~\bibnamefont
			{Sakaguchi}}\ and\ \bibinfo {author} {\bibfnamefont {Y.}~\bibnamefont
			{Kuramoto}},\ }\bibfield  {title} {\bibinfo {title} {A {Soluble} {Active}
			{Rotater} {Model} {Showing} {Phase} {Transitions} via {Mutual}
			{Entertainment}},\ }\href {https://doi.org/10.1143/PTP.76.576} {\bibfield
		{journal} {\bibinfo  {journal} {Progress of Theoretical Physics}\ }\textbf
		{\bibinfo {volume} {76}},\ \bibinfo {pages} {576} (\bibinfo {year}
		{1986})}\BibitemShut {NoStop}%
	\bibitem [{\citenamefont {Hong}\ and\ \citenamefont
		{Strogatz}(2011)}]{hong2011kuramoto}%
	\BibitemOpen
	\bibfield  {author} {\bibinfo {author} {\bibfnamefont {H.}~\bibnamefont
			{Hong}}\ and\ \bibinfo {author} {\bibfnamefont {S.~H.}\ \bibnamefont
			{Strogatz}},\ }\bibfield  {title} {\bibinfo {title} {Kuramoto model of
			coupled oscillators with positive and negative coupling parameters: an
			example of conformist and contrarian oscillators},\ }\href@noop {} {\bibfield
		{journal} {\bibinfo  {journal} {Physical Review Letters}\ }\textbf {\bibinfo
			{volume} {106}},\ \bibinfo {pages} {054102} (\bibinfo {year}
		{2011})}\BibitemShut {NoStop}%
	\bibitem [{\citenamefont {Yeung}\ and\ \citenamefont
		{Strogatz}(1999)}]{yeung1999time}%
	\BibitemOpen
	\bibfield  {author} {\bibinfo {author} {\bibfnamefont {M.~S.}\ \bibnamefont
			{Yeung}}\ and\ \bibinfo {author} {\bibfnamefont {S.~H.}\ \bibnamefont
			{Strogatz}},\ }\bibfield  {title} {\bibinfo {title} {Time delay in the
			kuramoto model of coupled oscillators},\ }\href@noop {} {\bibfield  {journal}
		{\bibinfo  {journal} {Physical Review Letters}\ }\textbf {\bibinfo {volume}
			{82}},\ \bibinfo {pages} {648} (\bibinfo {year} {1999})}\BibitemShut
	{NoStop}%
	\bibitem [{\citenamefont {Breakspear}\ \emph {et~al.}(2010)\citenamefont
		{Breakspear}, \citenamefont {Heitmann},\ and\ \citenamefont
		{Daffertshofer}}]{breakspear2010generative}%
	\BibitemOpen
	\bibfield  {author} {\bibinfo {author} {\bibfnamefont {M.}~\bibnamefont
			{Breakspear}}, \bibinfo {author} {\bibfnamefont {S.}~\bibnamefont
			{Heitmann}},\ and\ \bibinfo {author} {\bibfnamefont {A.}~\bibnamefont
			{Daffertshofer}},\ }\bibfield  {title} {\bibinfo {title} {Generative models
			of cortical oscillations: neurobiological implications of the kuramoto
			model},\ }\href@noop {} {\bibfield  {journal} {\bibinfo  {journal} {Frontiers
				in human neuroscience}\ }\textbf {\bibinfo {volume} {4}},\ \bibinfo {pages}
		{190} (\bibinfo {year} {2010})}\BibitemShut {NoStop}%
	\bibitem [{\citenamefont {Strogatz}(2001)}]{strogatz2001exploring}%
	\BibitemOpen
	\bibfield  {author} {\bibinfo {author} {\bibfnamefont {S.~H.}\ \bibnamefont
			{Strogatz}},\ }\bibfield  {title} {\bibinfo {title} {Exploring complex
			networks},\ }\href@noop {} {\bibfield  {journal} {\bibinfo  {journal}
			{Nature}\ }\textbf {\bibinfo {volume} {410}},\ \bibinfo {pages} {268}
		(\bibinfo {year} {2001})}\BibitemShut {NoStop}%
	\bibitem [{\citenamefont {Moreno}\ and\ \citenamefont
		{Pacheco}(2004)}]{moreno2004synchronization}%
	\BibitemOpen
	\bibfield  {author} {\bibinfo {author} {\bibfnamefont {Y.}~\bibnamefont
			{Moreno}}\ and\ \bibinfo {author} {\bibfnamefont {A.~F.}\ \bibnamefont
			{Pacheco}},\ }\bibfield  {title} {\bibinfo {title} {Synchronization of
			kuramoto oscillators in scale-free networks},\ }\href@noop {} {\bibfield
		{journal} {\bibinfo  {journal} {Europhysics Letters}\ }\textbf {\bibinfo
			{volume} {68}},\ \bibinfo {pages} {603} (\bibinfo {year} {2004})}\BibitemShut
	{NoStop}%
	\bibitem [{\citenamefont {Rodrigues}\ \emph {et~al.}(2016)\citenamefont
		{Rodrigues}, \citenamefont {Peron}, \citenamefont {Ji},\ and\ \citenamefont
		{Kurths}}]{Rodrigues2016}%
	\BibitemOpen
	\bibfield  {author} {\bibinfo {author} {\bibfnamefont {F.~A.}\ \bibnamefont
			{Rodrigues}}, \bibinfo {author} {\bibfnamefont {T.~K. D.~M.}\ \bibnamefont
			{Peron}}, \bibinfo {author} {\bibfnamefont {P.}~\bibnamefont {Ji}},\ and\
		\bibinfo {author} {\bibfnamefont {J.}~\bibnamefont {Kurths}},\ }\bibfield
	{title} {\bibinfo {title} {{The Kuramoto model in complex networks}},\ }\href
	{https://doi.org/10.1016/j.physrep.2015.10.008} {\bibfield  {journal}
		{\bibinfo  {journal} {Physics Reports}\ }\textbf {\bibinfo {volume} {610}},\
		\bibinfo {pages} {1} (\bibinfo {year} {2016})},\ \Eprint
	{https://arxiv.org/abs/1511.07139} {arXiv:1511.07139} \BibitemShut {NoStop}%
	\bibitem [{\citenamefont {Martens}\ \emph {et~al.}(2009)\citenamefont
		{Martens}, \citenamefont {Barreto}, \citenamefont {Strogatz}, \citenamefont
		{Ott}, \citenamefont {So},\ and\ \citenamefont
		{Antonsen}}]{martens2009exact}%
	\BibitemOpen
	\bibfield  {author} {\bibinfo {author} {\bibfnamefont {E.~A.}\ \bibnamefont
			{Martens}}, \bibinfo {author} {\bibfnamefont {E.}~\bibnamefont {Barreto}},
		\bibinfo {author} {\bibfnamefont {S.~H.}\ \bibnamefont {Strogatz}}, \bibinfo
		{author} {\bibfnamefont {E.}~\bibnamefont {Ott}}, \bibinfo {author}
		{\bibfnamefont {P.}~\bibnamefont {So}},\ and\ \bibinfo {author}
		{\bibfnamefont {T.~M.}\ \bibnamefont {Antonsen}},\ }\bibfield  {title}
	{\bibinfo {title} {Exact results for the kuramoto model with a bimodal
			frequency distribution},\ }\href@noop {} {\bibfield  {journal} {\bibinfo
			{journal} {Physical Review E}\ }\textbf {\bibinfo {volume} {79}},\ \bibinfo
		{pages} {026204} (\bibinfo {year} {2009})}\BibitemShut {NoStop}%
	\bibitem [{\citenamefont {Gomez-Gardenes}\ \emph {et~al.}(2011)\citenamefont
		{Gomez-Gardenes}, \citenamefont {Gomez}, \citenamefont {Arenas},\ and\
		\citenamefont {Moreno}}]{Gomez-Gardenes2011}%
	\BibitemOpen
	\bibfield  {author} {\bibinfo {author} {\bibfnamefont {J.}~\bibnamefont
			{Gomez-Gardenes}}, \bibinfo {author} {\bibfnamefont {S.}~\bibnamefont
			{Gomez}}, \bibinfo {author} {\bibfnamefont {A.}~\bibnamefont {Arenas}},\ and\
		\bibinfo {author} {\bibfnamefont {Y.}~\bibnamefont {Moreno}},\ }\bibfield
	{title} {\bibinfo {title} {{Explosive synchronization transitions in
				scale-free networks}},\ }\href
	{https://doi.org/10.1103/PhysRevLett.106.128701} {\bibfield  {journal}
		{\bibinfo  {journal} {Physical Review Letters}\ }\textbf {\bibinfo {volume}
			{106}},\ \bibinfo {pages} {1} (\bibinfo {year} {2011})},\ \Eprint
	{https://arxiv.org/abs/1102.4823} {arXiv:1102.4823} \BibitemShut {NoStop}%
	\bibitem [{\citenamefont {Ji}\ \emph {et~al.}(2013)\citenamefont {Ji},
		\citenamefont {Peron}, \citenamefont {Menck}, \citenamefont {Rodrigues},\
		and\ \citenamefont {Kurths}}]{Ji2013}%
	\BibitemOpen
	\bibfield  {author} {\bibinfo {author} {\bibfnamefont {P.}~\bibnamefont
			{Ji}}, \bibinfo {author} {\bibfnamefont {T.~K.~D.}\ \bibnamefont {Peron}},
		\bibinfo {author} {\bibfnamefont {P.~J.}\ \bibnamefont {Menck}}, \bibinfo
		{author} {\bibfnamefont {F.~A.}\ \bibnamefont {Rodrigues}},\ and\ \bibinfo
		{author} {\bibfnamefont {J.}~\bibnamefont {Kurths}},\ }\bibfield  {title}
	{\bibinfo {title} {{Cluster explosive synchronization in complex networks}},\
	}\href {https://doi.org/10.1103/PhysRevLett.110.218701} {\bibfield  {journal}
		{\bibinfo  {journal} {Physical Review Letters}\ }\textbf {\bibinfo {volume}
			{110}},\ \bibinfo {pages} {1} (\bibinfo {year} {2013})},\ \Eprint
	{https://arxiv.org/abs/arXiv:1303.3498v2} {arXiv:arXiv:1303.3498v2}
	\BibitemShut {NoStop}%
	\bibitem [{\citenamefont {Acebr{\'{o}}n}\ \emph {et~al.}(2005)\citenamefont
		{Acebr{\'{o}}n}, \citenamefont {Bonilla}, \citenamefont {Vicente},
		\citenamefont {Ritort},\ and\ \citenamefont {Spigler}}]{Acebron2005}%
	\BibitemOpen
	\bibfield  {author} {\bibinfo {author} {\bibfnamefont {J.~A.}\ \bibnamefont
			{Acebr{\'{o}}n}}, \bibinfo {author} {\bibfnamefont {L.~L.}\ \bibnamefont
			{Bonilla}}, \bibinfo {author} {\bibfnamefont {C.~J.~P.}\ \bibnamefont
			{Vicente}}, \bibinfo {author} {\bibfnamefont {F.}~\bibnamefont {Ritort}},\
		and\ \bibinfo {author} {\bibfnamefont {R.}~\bibnamefont {Spigler}},\
	}\bibfield  {title} {\bibinfo {title} {{The Kuramoto model: A simple paradigm
				for synchronization phenomena}},\ }\href
	{https://doi.org/10.1103/RevModPhys.77.137} {\bibfield  {journal} {\bibinfo
			{journal} {Reviews of Modern Physics}\ }\textbf {\bibinfo {volume} {77}},\
		\bibinfo {pages} {137} (\bibinfo {year} {2005})},\ \Eprint
	{https://arxiv.org/abs/0306625} {arXiv:0306625 [cond-mat]} \BibitemShut
	{NoStop}%
	\bibitem [{\citenamefont {D{\"o}rfler}\ and\ \citenamefont
		{Bullo}(2011)}]{dorfler2011critical}%
	\BibitemOpen
	\bibfield  {author} {\bibinfo {author} {\bibfnamefont {F.}~\bibnamefont
			{D{\"o}rfler}}\ and\ \bibinfo {author} {\bibfnamefont {F.}~\bibnamefont
			{Bullo}},\ }\bibfield  {title} {\bibinfo {title} {On the critical coupling
			for kuramoto oscillators},\ }\href@noop {} {\bibfield  {journal} {\bibinfo
			{journal} {SIAM Journal on Applied Dynamical Systems}\ }\textbf {\bibinfo
			{volume} {10}},\ \bibinfo {pages} {1070} (\bibinfo {year}
		{2011})}\BibitemShut {NoStop}%
	\bibitem [{\citenamefont {Olmi}\ \emph {et~al.}(2014)\citenamefont {Olmi},
		\citenamefont {Navas}, \citenamefont {Boccaletti},\ and\ \citenamefont
		{Torcini}}]{olmi2014hysteretic}%
	\BibitemOpen
	\bibfield  {author} {\bibinfo {author} {\bibfnamefont {S.}~\bibnamefont
			{Olmi}}, \bibinfo {author} {\bibfnamefont {A.}~\bibnamefont {Navas}},
		\bibinfo {author} {\bibfnamefont {S.}~\bibnamefont {Boccaletti}},\ and\
		\bibinfo {author} {\bibfnamefont {A.}~\bibnamefont {Torcini}},\ }\bibfield
	{title} {\bibinfo {title} {Hysteretic transitions in the kuramoto model with
			inertia},\ }\href@noop {} {\bibfield  {journal} {\bibinfo  {journal}
			{Physical Review E}\ }\textbf {\bibinfo {volume} {90}},\ \bibinfo {pages}
		{042905} (\bibinfo {year} {2014})}\BibitemShut {NoStop}%
	\bibitem [{\citenamefont {Childs}\ and\ \citenamefont
		{Strogatz}(2008)}]{Childs2008}%
	\BibitemOpen
	\bibfield  {author} {\bibinfo {author} {\bibfnamefont {L.~M.}\ \bibnamefont
			{Childs}}\ and\ \bibinfo {author} {\bibfnamefont {S.~H.}\ \bibnamefont
			{Strogatz}},\ }\bibfield  {title} {\bibinfo {title} {Stability diagram for
			the forced kuramoto model},\ }\href {https://doi.org/10.1063/1.3049136}
	{\bibfield  {journal} {\bibinfo  {journal} {Chaos: An Interdisciplinary
				Journal of Nonlinear Science}\ }\textbf {\bibinfo {volume} {18}},\ \bibinfo
		{pages} {043128} (\bibinfo {year} {2008})}\BibitemShut {NoStop}%
	\bibitem [{\citenamefont {Moreira}\ and\ \citenamefont {{de
				Aguiar}}(2019)}]{moreira2019}%
	\BibitemOpen
	\bibfield  {author} {\bibinfo {author} {\bibfnamefont {C.~A.}\ \bibnamefont
			{Moreira}}\ and\ \bibinfo {author} {\bibfnamefont {M.~A.}\ \bibnamefont {{de
					Aguiar}}},\ }\bibfield  {title} {\bibinfo {title} {Global synchronization of
			partially forced kuramoto oscillators on networks},\ }\href
	{https://doi.org/https://doi.org/10.1016/j.physa.2018.09.096} {\bibfield
		{journal} {\bibinfo  {journal} {Physica A: Statistical Mechanics and its
				Applications}\ }\textbf {\bibinfo {volume} {514}},\ \bibinfo {pages} {487}
		(\bibinfo {year} {2019})}\BibitemShut {NoStop}%
	\bibitem [{\citenamefont {Chandra}\ \emph
		{et~al.}(2019{\natexlab{a}})\citenamefont {Chandra}, \citenamefont {Girvan},\
		and\ \citenamefont {Ott}}]{chandra2019continuous}%
	\BibitemOpen
	\bibfield  {author} {\bibinfo {author} {\bibfnamefont {S.}~\bibnamefont
			{Chandra}}, \bibinfo {author} {\bibfnamefont {M.}~\bibnamefont {Girvan}},\
		and\ \bibinfo {author} {\bibfnamefont {E.}~\bibnamefont {Ott}},\ }\bibfield
	{title} {\bibinfo {title} {Continuous versus discontinuous transitions in the
			d-dimensional generalized kuramoto model: Odd d is different},\ }\href@noop
	{} {\bibfield  {journal} {\bibinfo  {journal} {Physical Review X}\ }\textbf
		{\bibinfo {volume} {9}},\ \bibinfo {pages} {011002} (\bibinfo {year}
		{2019}{\natexlab{a}})}\BibitemShut {NoStop}%
	\bibitem [{\citenamefont {Barioni}\ and\ \citenamefont
		{de~Aguiar}(2021)}]{barioni2021}%
	\BibitemOpen
	\bibfield  {author} {\bibinfo {author} {\bibfnamefont {A.~E.~D.}\
			\bibnamefont {Barioni}}\ and\ \bibinfo {author} {\bibfnamefont {M.~A.~M.}\
			\bibnamefont {de~Aguiar}},\ }\bibfield  {title} {\bibinfo {title} {Complexity
			reduction in the {3D} {Kuramoto} model},\ }\href
	{https://doi.org/10.1016/j.chaos.2021.111090} {\bibfield  {journal} {\bibinfo
			{journal} {Chaos, Solitons \& Fractals}\ }\textbf {\bibinfo {volume}
			{149}},\ \bibinfo {pages} {111090} (\bibinfo {year} {2021})}\BibitemShut
	{NoStop}%
	\bibitem [{\citenamefont {de~Aguiar}(2023)}]{de2023generalized}%
	\BibitemOpen
	\bibfield  {author} {\bibinfo {author} {\bibfnamefont {M.~A.}\ \bibnamefont
			{de~Aguiar}},\ }\bibfield  {title} {\bibinfo {title} {Generalized frustration
			in the multidimensional kuramoto model},\ }\href@noop {} {\bibfield
		{journal} {\bibinfo  {journal} {Physical Review E}\ }\textbf {\bibinfo
			{volume} {107}},\ \bibinfo {pages} {044205} (\bibinfo {year}
		{2023})}\BibitemShut {NoStop}%
	\bibitem [{\citenamefont {Battiston}\ \emph {et~al.}(2020)\citenamefont
		{Battiston}, \citenamefont {Cencetti}, \citenamefont {Iacopini},
		\citenamefont {Latora}, \citenamefont {Lucas}, \citenamefont {Patania},
		\citenamefont {Young},\ and\ \citenamefont {Petri}}]{battiston2020networks}%
	\BibitemOpen
	\bibfield  {author} {\bibinfo {author} {\bibfnamefont {F.}~\bibnamefont
			{Battiston}}, \bibinfo {author} {\bibfnamefont {G.}~\bibnamefont {Cencetti}},
		\bibinfo {author} {\bibfnamefont {I.}~\bibnamefont {Iacopini}}, \bibinfo
		{author} {\bibfnamefont {V.}~\bibnamefont {Latora}}, \bibinfo {author}
		{\bibfnamefont {M.}~\bibnamefont {Lucas}}, \bibinfo {author} {\bibfnamefont
			{A.}~\bibnamefont {Patania}}, \bibinfo {author} {\bibfnamefont {J.-G.}\
			\bibnamefont {Young}},\ and\ \bibinfo {author} {\bibfnamefont
			{G.}~\bibnamefont {Petri}},\ }\bibfield  {title} {\bibinfo {title} {Networks
			beyond pairwise interactions: Structure and dynamics},\ }\href@noop {}
	{\bibfield  {journal} {\bibinfo  {journal} {Physics Reports}\ }\textbf
		{\bibinfo {volume} {874}},\ \bibinfo {pages} {1} (\bibinfo {year}
		{2020})}\BibitemShut {NoStop}%
	\bibitem [{\citenamefont {Muolo}\ \emph {et~al.}(2024)\citenamefont {Muolo},
		\citenamefont {Njougouo}, \citenamefont {Gambuzza}, \citenamefont
		{Carletti},\ and\ \citenamefont {Frasca}}]{muolo2024phase}%
	\BibitemOpen
	\bibfield  {author} {\bibinfo {author} {\bibfnamefont {R.}~\bibnamefont
			{Muolo}}, \bibinfo {author} {\bibfnamefont {T.}~\bibnamefont {Njougouo}},
		\bibinfo {author} {\bibfnamefont {L.~V.}\ \bibnamefont {Gambuzza}}, \bibinfo
		{author} {\bibfnamefont {T.}~\bibnamefont {Carletti}},\ and\ \bibinfo
		{author} {\bibfnamefont {M.}~\bibnamefont {Frasca}},\ }\bibfield  {title}
	{\bibinfo {title} {Phase chimera states on nonlocal hyperrings},\ }\href@noop
	{} {\bibfield  {journal} {\bibinfo  {journal} {Physical Review E}\ }\textbf
		{\bibinfo {volume} {109}},\ \bibinfo {pages} {L022201} (\bibinfo {year}
		{2024})}\BibitemShut {NoStop}%
	\bibitem [{\citenamefont {O'Keeffe}\ \emph {et~al.}(2017)\citenamefont
		{O'Keeffe}, \citenamefont {Hong},\ and\ \citenamefont
		{Strogatz}}]{o2017oscillators}%
	\BibitemOpen
	\bibfield  {author} {\bibinfo {author} {\bibfnamefont {K.~P.}\ \bibnamefont
			{O'Keeffe}}, \bibinfo {author} {\bibfnamefont {H.}~\bibnamefont {Hong}},\
		and\ \bibinfo {author} {\bibfnamefont {S.~H.}\ \bibnamefont {Strogatz}},\
	}\bibfield  {title} {\bibinfo {title} {Oscillators that sync and swarm},\
	}\href@noop {} {\bibfield  {journal} {\bibinfo  {journal} {Nature
				Communications}\ }\textbf {\bibinfo {volume} {8}},\ \bibinfo {pages} {1504}
		(\bibinfo {year} {2017})}\BibitemShut {NoStop}%
	\bibitem [{\citenamefont {Sar}\ \emph {et~al.}(2026)\citenamefont {Sar},
		\citenamefont {O'Keeffe}, \citenamefont {Liz{\'a}rraga}, \citenamefont
		{de~Aguiar}, \citenamefont {Bettstetter},\ and\ \citenamefont
		{Ghosh}}]{sar2026interplay}%
	\BibitemOpen
	\bibfield  {author} {\bibinfo {author} {\bibfnamefont {G.~K.}\ \bibnamefont
			{Sar}}, \bibinfo {author} {\bibfnamefont {K.}~\bibnamefont {O'Keeffe}},
		\bibinfo {author} {\bibfnamefont {J.~U.}\ \bibnamefont {Liz{\'a}rraga}},
		\bibinfo {author} {\bibfnamefont {M.~A.}\ \bibnamefont {de~Aguiar}}, \bibinfo
		{author} {\bibfnamefont {C.}~\bibnamefont {Bettstetter}},\ and\ \bibinfo
		{author} {\bibfnamefont {D.}~\bibnamefont {Ghosh}},\ }\bibfield  {title}
	{\bibinfo {title} {Interplay of sync and swarm: Theory and application of
			swarmalators},\ }\href@noop {} {\bibfield  {journal} {\bibinfo  {journal}
			{Physics Reports}\ }\textbf {\bibinfo {volume} {1167}},\ \bibinfo {pages} {1}
		(\bibinfo {year} {2026})}\BibitemShut {NoStop}%
	\bibitem [{\citenamefont {Buzanello}\ \emph {et~al.}(2022)\citenamefont
		{Buzanello}, \citenamefont {Barioni},\ and\ \citenamefont
		{de~Aguiar}}]{Buzanello2022}%
	\BibitemOpen
	\bibfield  {author} {\bibinfo {author} {\bibfnamefont {G.~L.}\ \bibnamefont
			{Buzanello}}, \bibinfo {author} {\bibfnamefont {A.~E.~D.}\ \bibnamefont
			{Barioni}},\ and\ \bibinfo {author} {\bibfnamefont {M.~A.~M.}\ \bibnamefont
			{de~Aguiar}},\ }\bibfield  {title} {\bibinfo {title} {Matrix coupling and
			generalized frustration in kuramoto oscillators},\ }\href
	{https://doi.org/10.1063/5.0108672} {\bibfield  {journal} {\bibinfo
			{journal} {Chaos: An Interdisciplinary Journal of Nonlinear Science}\
		}\textbf {\bibinfo {volume} {32}},\ \bibinfo {pages} {093130} (\bibinfo
		{year} {2022})}\BibitemShut {NoStop}%
	\bibitem [{\citenamefont {Ott}\ and\ \citenamefont {Antonsen}(2008)}]{Ott2008}%
	\BibitemOpen
	\bibfield  {author} {\bibinfo {author} {\bibfnamefont {E.}~\bibnamefont
			{Ott}}\ and\ \bibinfo {author} {\bibfnamefont {T.~M.}\ \bibnamefont
			{Antonsen}},\ }\bibfield  {title} {\bibinfo {title} {{Low dimensional
				behavior of large systems of globally coupled oscillators}},\ }\href
	{https://doi.org/10.1063/1.2930766} {\bibfield  {journal} {\bibinfo
			{journal} {Chaos: An Interdisciplinary Journal of Nonlinear Science}\
		}\textbf {\bibinfo {volume} {18}},\ \bibinfo {pages} {037113} (\bibinfo
		{year} {2008})}\BibitemShut {NoStop}%
	\bibitem [{\citenamefont {Givental}\ \emph {et~al.}(2009)\citenamefont
		{Givental}, \citenamefont {Khesin}, \citenamefont {Marsden}, \citenamefont
		{Varchenko}, \citenamefont {Vassiliev}, \citenamefont {Viro},\ and\
		\citenamefont {Zakalyukin}}]{Arnold2009}%
	\BibitemOpen
	\bibinfo {editor} {\bibfnamefont {A.~B.}\ \bibnamefont {Givental}}, \bibinfo
	{editor} {\bibfnamefont {B.~A.}\ \bibnamefont {Khesin}}, \bibinfo {editor}
	{\bibfnamefont {J.~E.}\ \bibnamefont {Marsden}}, \bibinfo {editor}
	{\bibfnamefont {A.~N.}\ \bibnamefont {Varchenko}}, \bibinfo {editor}
	{\bibfnamefont {V.~A.}\ \bibnamefont {Vassiliev}}, \bibinfo {editor}
	{\bibfnamefont {O.~Y.}\ \bibnamefont {Viro}},\ and\ \bibinfo {editor}
	{\bibfnamefont {V.~M.}\ \bibnamefont {Zakalyukin}},\ eds.,\ \bibinfo {title}
	{Small denominators. i. mapping of the circumference onto itself},\ in\ \href
	{https://doi.org/10.1007/978-3-642-01742-1_10} {\emph {\bibinfo {booktitle}
			{Collected Works: Representations of Functions, Celestial Mechanics and KAM
				Theory, 1957--1965}}}\ (\bibinfo  {publisher} {Springer Berlin Heidelberg},\
	\bibinfo {address} {Berlin, Heidelberg},\ \bibinfo {year} {2009})\ pp.\
	\bibinfo {pages} {152--223}\BibitemShut {NoStop}%
	\bibitem [{\citenamefont {Chandra}\ \emph
		{et~al.}(2019{\natexlab{b}})\citenamefont {Chandra}, \citenamefont {Girvan},\
		and\ \citenamefont {Ott}}]{chandra2019}%
	\BibitemOpen
	\bibfield  {author} {\bibinfo {author} {\bibfnamefont {S.}~\bibnamefont
			{Chandra}}, \bibinfo {author} {\bibfnamefont {M.}~\bibnamefont {Girvan}},\
		and\ \bibinfo {author} {\bibfnamefont {E.}~\bibnamefont {Ott}},\ }\bibfield
	{title} {\bibinfo {title} {Continuous versus discontinuous transitions in the
			$d$-dimensional generalized kuramoto model: Odd $d$ is different},\ }\href
	{https://doi.org/10.1103/PhysRevX.9.011002} {\bibfield  {journal} {\bibinfo
			{journal} {Phys. Rev. X}\ }\textbf {\bibinfo {volume} {9}},\ \bibinfo {pages}
		{011002} (\bibinfo {year} {2019}{\natexlab{b}})}\BibitemShut {NoStop}%
	\bibitem [{\citenamefont {Costa}\ and\ \citenamefont
		{de~Aguiar}(2024)}]{costa2024dynamics}%
	\BibitemOpen
	\bibfield  {author} {\bibinfo {author} {\bibfnamefont {G.~S.}\ \bibnamefont
			{Costa}}\ and\ \bibinfo {author} {\bibfnamefont {M.~A.}\ \bibnamefont
			{de~Aguiar}},\ }\bibfield  {title} {\bibinfo {title} {Dynamics of matrix
			coupled kuramoto oscillators on modular networks: excitable behavior and
			global decoherence},\ }\href@noop {} {\bibfield  {journal} {\bibinfo
			{journal} {arXiv preprint arXiv:2403.06689}\ } (\bibinfo {year}
		{2024})}\BibitemShut {NoStop}%
	\bibitem [{\citenamefont {Sakaguchi}\ \emph {et~al.}(1988)\citenamefont
		{Sakaguchi}, \citenamefont {Shinomoto},\ and\ \citenamefont
		{Kuramoto}}]{Sakaguchi1988}%
	\BibitemOpen
	\bibfield  {author} {\bibinfo {author} {\bibfnamefont {H.}~\bibnamefont
			{Sakaguchi}}, \bibinfo {author} {\bibfnamefont {S.}~\bibnamefont
			{Shinomoto}},\ and\ \bibinfo {author} {\bibfnamefont {Y.}~\bibnamefont
			{Kuramoto}},\ }\bibfield  {title} {\bibinfo {title} {{Phase Transitions and
				Their Bifurcation Analysis in a Large Population of Active Rotators with
				Mean-Field Coupling}},\ }\href {https://doi.org/10.1143/PTP.79.600}
	{\bibfield  {journal} {\bibinfo  {journal} {Progress of Theoretical Physics}\
		}\textbf {\bibinfo {volume} {79}},\ \bibinfo {pages} {600} (\bibinfo {year}
		{1988})}\BibitemShut {NoStop}%
	\bibitem [{\citenamefont {Escande}(1985)}]{escande1985stochasticity}%
	\BibitemOpen
	\bibfield  {author} {\bibinfo {author} {\bibfnamefont {D.~F.}\ \bibnamefont
			{Escande}},\ }\bibfield  {title} {\bibinfo {title} {Stochasticity in
			classical hamiltonian systems: universal aspects},\ }\href@noop {} {\bibfield
		{journal} {\bibinfo  {journal} {Physics Reports}\ }\textbf {\bibinfo
			{volume} {121}},\ \bibinfo {pages} {165} (\bibinfo {year}
		{1985})}\BibitemShut {NoStop}%
	\bibitem [{\citenamefont {Odor}\ and\ \citenamefont {Deng}(2023)}]{odor2023}%
	\BibitemOpen
	\bibfield  {author} {\bibinfo {author} {\bibfnamefont {G.}~\bibnamefont
			{Odor}}\ and\ \bibinfo {author} {\bibfnamefont {S.}~\bibnamefont {Deng}},\
	}\bibfield  {title} {\bibinfo {title} {Synchronization transition of the
			second-order kuramoto model on lattices},\ }\bibfield  {journal} {\bibinfo
		{journal} {Entropy}\ }\textbf {\bibinfo {volume} {25}},\ \href
	{https://doi.org/10.3390/e25010164} {10.3390/e25010164} (\bibinfo {year}
	{2023})\BibitemShut {NoStop}%
\end{thebibliography}

%

\end{document}